\documentclass{article}

\usepackage{arxiv}

\usepackage[utf8]{inputenc} 
\usepackage[T1]{fontenc}    
\usepackage{hyperref}       
\usepackage{url}            
\usepackage{booktabs}       
\usepackage{amsfonts}       
\usepackage{nicefrac}       
\usepackage{microtype}      
\usepackage{lipsum}		
\usepackage{graphicx}
\usepackage{doi}

\usepackage{amsmath,amsthm,amssymb,mathtools,mathrsfs,amsfonts}

\def\bbb[#1]{\boldsymbol{#1}}
\def\mmm[#1]{\mathcal{#1}}
\def\sss[#1]{\mathscr{#1}}

\title{Ensemble-Based Experimental Design for Targeting Data Acquisition to Inform Climate Models}

\author{Oliver R. A. Dunbar \\
	Department of Environmental Science and Engineering,\\
	California Institute of Technology,\\
	Pasadena, CA, USA.\\
	\texttt{odunbar@caltech.edu}\\
	\And 
	Michael F. Howland \\
	Civil and Environmental Engineering,\\
	Massachusetts Institute of Technology,\\
	Cambridge, MA, USA.\\
	\And
	Tapio Schneider \\
	Department of Environmental Science and Engineering,\\
	California Institute of Technology,\\
	Pasadena, CA, USA.\\
	\And
	Andrew M. Stuart \\
	Department of Computing and Mathematical Sciences,\\
	California Institute of Technology,\\
	Pasadena, CA, USA.\\
}

\renewcommand{\shorttitle}{}

\hypersetup{
pdftitle={Ensemble-Based Experimental Design for Targeting Data Acquisition to Inform Climate Models},
pdfsubject={},
pdfauthor={},
pdfkeywords={},
}

\begin{document}
\maketitle

\begin{abstract}
Data required to calibrate uncertain GCM parameterizations are often only available in limited regions or time periods, for example, observational data from field campaigns, or data generated in local high-resolution simulations. This raises the question of where and when to acquire additional data to be maximally informative about parameterizations in a GCM. Here we construct a new ensemble-based parallel algorithm to automatically target data acquisition to regions and times that maximize the uncertainty reduction, or information gain, about GCM parameters. The algorithm uses a Bayesian framework that exploits a quantified distribution of GCM parameters as a measure of uncertainty. This distribution is informed by time-averaged climate statistics restricted to local regions and times. The algorithm is embedded in the recently developed calibrate-emulate-sample (CES) framework, which performs efficient model calibration and uncertainty quantification with only $\mmm[O](10^2)$ model evaluations, compared with $\mmm[O](10^5)$ evaluations typically needed for traditional approaches to Bayesian calibration. We demonstrate the algorithm with an idealized GCM, with which we generate surrogates of local data. In this perfect-model setting, we calibrate parameters and quantify uncertainties in a quasi-equilibrium convection scheme in the GCM. We consider targeted data that are (i) localized in space for statistically stationary simulations, and (ii) localized in space and time for seasonally varying simulations. In these proof-of-concept applications, the calculated information gain reflects the reduction in parametric uncertainty obtained from Bayesian inference when harnessing a targeted sample of data. The largest information gain typically, but not always, results from regions near the intertropical convergence zone (ITCZ).
\end{abstract}


\section{Introduction}

Parameterizations of subgrid-scale processes, such as the turbulence and convection controlling clouds, are the principal cause of physical uncertainties in climate predictions \cite{Ces-etal89,Cess90a,BonDuf05,Stephens05,Bony06,Vial13a,Webb13b,Brient16b}. Such parametric uncertainties in principle can be quantified and reduced by calibration with data. High-resolution simulations such as large-eddy simulations (LES) are able to resolve turbulence and convection in atmosphere and oceans over limited areas \cite{Siebesma03,Stevens05a,Khairoutdinov09a,Matheou14a,Schalkwijk15a,Pressel15a,Pressel17a} and have been used to calibrate climate model parameterizations at selected sites 
e.g., \cite{GCSS93,Liu01a,Siebesma03,Siebesma07,Hohenegger11a,Zhang13a,de-Rooy13a,Romps16a,Sou-etal20,FoxLi17,Tan18a,Smalley19a,Couvreux21a,Hourdin21b}. More systematically, one can drive LES with a coarse-resolution general circulation model (GCM) \cite{Shen20a,JarSchSheSriZhi21-pre}, giving the freedom to run LES at many sites across the globe, at different time periods in the seasonal cycle, and in changed climates, with a more consistent forcing scenario than in previous idealized setups.

A natural question arises: how might we most effectively place such high-resolution simulations? In this paper we address the general task of optimal targeting of data acquisition, and we demonstrate our approach within an idealized GCM setting. We create an automated algorithm based on experimental design criteria \cite{ChaVer95} to choose data acquisition sites and time periods that are maximally informative about parameters in a model. The experimental-design problem we address has similarities with the problem of how to choose sites for targeted weather observations to optimally improve weather forecasts \cite{Lorenz98a,bishop1999ensemble, emanuel1995report}. However, in contrast to the situation in weather forecasting, which focuses on trajectory matching for state estimation, our focus is on minimizing mismatches in time-averaged climate statistics for the estimation of parameters in climate models.  

To learn from time-averaged statistics, we adopt a Bayesian inverse problem setting (see, e.g., \cite{book:KaiSom06}, \cite{book:Tar05}, \cite{Stu10}, and \cite{pre:DasStu13} for reviews). In this setting, parameters (or parametric or nonparametric functions) in parameterizations are treated as having probability distributions. Data (e.g., climate statistics) are used to reduce the uncertainty reflected by these distributions, balancing contributions of the data with that of prior knowledge about parameters (e.g., physical constraints). This results in the joint posterior distribution for parameters, including the correlation structure of uncertainties among parameters. The Bayesian experimental design tools we apply in this paper leverage the posterior distribution to determine regions and times where local data are maximally effective at reducing parameter uncertainties. As is typical in such analyses, we measure the quality of a design (site location or time period) by a scalar utility function. We choose a utility that quantifies the information entropy loss between posterior and prior for each design e.g., \cite{ChaVer95,book:FedHac97,DroMcGPet13}. The site and time period of maximal utility determines where to acquire data. 

Construction of the joint posterior distribution of the parameters is well known to be a computationally intensive task, with commonly used Markov chain Monte Carlo (MCMC) methods typically requiring $\mathcal{O}(10^5)$ evaluations of the model in which the parameters appear (see \cite{Gey11} for an overview). The recent development of the calibrate-emulate-sample (CES) framework accelerates Bayesian learning by a factor of $10^3$ \cite{CleGarLanSchStu21,Dunbar21a}. The calibration stage uses a variant of ensemble Kalman inversion \cite{IglLawStu13,CheOli12,EmeRey13,Rei11} to obtain a collection of samples of the model about an optimal set of parameters. The emulation stage features the training of a machine learning emulator, here, a Gaussian process \cite{book:RasWil06,KenOHa00,Kennedy01a}, to emulate output statistics of the model using the pairs of parameters and model outputs from the calibration stage. The sample stage then samples a posterior distribution with MCMC methods, replacing the computationally expensive model with the cheap emulator. This framework can extend to the learning of data-driven parameterizations or other non-parametric functions, such as structural model errors e.g., \cite{LevStu21-pre,Schneider22v,Lopez-Gomez22a}. Our proposed algorithm builds on CES to incorporate Bayesian experimental design at negligible additional computational expense. In particular, we do not require additional forward model (GCM) evaluations over what is already required in CES to perform uncertainty quantification.

We demonstrate our approach with an idealized moist GCM \cite{Frierson06a} with modifications introduced by \cite{OGorman08b}, with which we generate surrogates of local data and in which we calibrate parameters in a quasi-equilibrium convection scheme \cite{Frierson07b}. We conduct numerical experiments with the idealized GCM in statistically stationary and seasonally varying configurations and show how to determine the utility of data at different sites and in different seasons. These experiments serve as proof-of-concept of the broad-purpose algorithm, which can be applied, for example, to determine optimal sites and times for high-resolution simulations for the calibration and uncertainty quantification of parameterizations.

In Section \ref{sec:method}, we define the inverse problems for parameter calibration and the optimal design algorithm; details of efficient uncertainty quantification (CES) are provided in Appendix \ref{sec:CES}. In Section \ref{sec:GCM}, we briefly describe the GCM used for demonstrating the algorithm. Results of the optimal design algorithm are described in Section \ref{s:results}. We end with a summary and discussion of conclusions in Section \ref{sec:conclusion}.

\section{Methodology}\label{sec:method}

Our goal is to target data acquisition to regions and times at which uncertainty reduction (information gain) is maximized. We do this in two stages. First, we learn the temporally and spatially varying sensitivities of the model statistics with respect to model parameters. Second, we use this knowledge to target data acquisition to regions and times at which the model is maximally sensitive to new data. We work in a framework similar to \cite{Dunbar21a} which focuses on accelerated uncertainty quantification within a GCM.

\subsection{Inverse problem to learn from limited-area data}\label{sec:prob}

We study calibration of parameters in a GCM by formulating parameter learning as a Bayesian inverse problem. Define $\mmm[G]_T(\bbb[\theta];\bbb[v]^{(0)})$ to be the forward map sending the parameters $\bbb[\theta]$ to time-aggregated simulated climate statistics (averaged over a window of length $T>0$) from an initial state $\bbb[v]^{(0)}$. We assume that the aggregation $\mmm[G]_T(\bbb[\theta],\cdot)$ is statistically stationary, and we refer to samples of such aggregated climate statistics as data throughout, irrespective of whether they are observational or computationally generated. We consider a situation in which data are only locally available, at a particular spatial or spatio-temporal location, indexed by $k$, which we refer to as the design point. We make use of a restriction operation $W_k$ to a point $k$, and define the limited-area forward map, $\mmm[S]_T(\bbb[\theta];k,\bbb[v]^{(0)}) = W_k\mmm[G]_T(\bbb[\theta];\bbb[v]^{(0)})$. 

For any given $k$, assume we have data $\bbb[z]_k$ available. For example, $\bbb[z]_k$ could be produced with a simulation with limited spatial or temporal extent, or by running a field campaign. 
We form an inverse problem for GCM learning from this data as
\begin{equation}\label{eq:localIPT}
 \bbb[z]_k = \mmm[S]_{T}(\bbb[\theta];k,\bbb[v]^{(0)})+ \delta_k,
\end{equation} 
where $\delta_k$ is a stochastic term to capture discrepancies between model $\mmm[S]_T(\cdot;k,\cdot)$ and data $\bbb[z]_k$, e.g., \cite{Kennedy01a}. The initial condition $\bbb[v]^{(0)}$ appears in this formulation but is treated as a nuisance variable. This view is justified in the context of learning about atmospheric parameterizations for climate models, where lower-frequency information (e.g., seasonal variations) is particular informative \cite{Schneider17c}. Indeed, the time-averaged data filters out the high-frequency information. Following \cite{Dunbar21a}, we write $\mmm[S]_T(\bbb[\theta];k,\bbb[v]^{(0)}) \approx  \mmm[S]_\infty(\bbb[\theta];k) + \sigma_k,$ where $\sigma_k\sim N(0, \Sigma(\bbb[\theta]))$ is normal noise, independent from $\delta_k$, with mean zero and with a covariance matrix $\Sigma(\bbb[\theta])$ reflecting chaotic internal variability. The Gaussian assumption is justified on the basis of a central limit theorem (CLT) applied to the time averages. The inverse problem \eqref{eq:localIPT} is thus approximated by the problem
\begin{equation}\label{eq:localIPinfa}
 \bbb[z]_k = \mmm[S]_{\infty}(\bbb[\theta];k) + \delta_k + \sigma_k,\qquad \sigma \sim N(0,W_k\Sigma(\bbb[\theta]) W_k^T).
\end{equation}
This is a desirable inverse problem without dependence on the initial condition. It is an approximation due to the CLT, but this approximation should be suitable if $T$ is taken larger than the dynamical system's Lyapunov timescale (for the atmosphere, this equates to $T\gtrsim 15$ days \cite{BuiCheEmaMagSunZha19}). In our experiments, we take $T=90$ days (Section~\ref{sec:obj_func}), or $T=30$ days (Appendix \ref{sec:30day}).  

Solving \eqref{eq:localIPinfa} involves finding the posterior distribution of $\bbb[\theta]$ given the data $\bbb[z]_k$, denoted $(\bbb[\theta] \mid \bbb[z]_k)$. Although we cannot evaluate $\mmm[S]_\infty$ directly, the emulate phase of the calibrate-emulate-sample (CES) algorithm \cite{CleGarLanSchStu21} constructs a surrogate of $\mmm[S]_\infty$ from carefully chosen evaluations of $\mmm[S]_T$. Details of the algorithm are provided in Appendix \ref{sec:CES}.

\subsection{Experimental design}\label{sec:des}
We consider a situation in which acquiring $\bbb[z]_k$ is associated with large costs. For example, $\bbb[z]_k$ could be data obtained by running a computationally demanding simulation, or running an expensive field campaign. Our starting point is to assume that a limited budget restricts us to evaluate $\bbb[z]_k$ at a single design point $k$ at a time. We want to choose the design point $k$ that leads to the most informative inverse problem \eqref{eq:localIPinfa}. We continue using a Bayesian point of view, namely, the optimal $k$ is the one for which the posterior distribution of $(\bbb[\theta]\mid\bbb[z]_k)$ learned from the inverse problem \eqref{eq:localIPinfa} has the smallest uncertainty. 
This perspective is motivated by the downstream goal of minimizing the parametric uncertainty of GCM predictions.

To answer this conclusively, one would need to evaluate $\bbb[z]_k$ at all design points $k$, which here is too computationally expensive. Instead, we investigate only the sensitivity of the forward model statistics $\mmm[G]_T$ to its parameters $\bbb[\theta]$ to assess the additional information provided at each design point $k$. This additional information at $k$ is used as a proxy for the information content that would exist when learning from data $\bbb[z]_k$. The benefits of this approach are that (i) we do not require any evaluations of $\bbb[z]_k$ to select the optimal location; (ii) the measure of information content is naturally constructed from the uncertainty reflected by the Bayesian posterior distribution; and (iii) we can perform this efficiently, and in a embarrassingly parallel fashion, requiring only $O(100)$ GCM runs, determined by the product of the ensemble size and the number of iterations typically needed in the calibration stage of the CES algorithm (see Appendix \ref{sec:CES}). The approach necessarily will contain a bias from the prior distribution of the parameters.


Each evaluation of the forward map involves a simulation with the GCM and thus depends on an initial condition $\bbb[v]^{(0)}$ and parameters $\bbb[\theta]$. Together this gives rise to the definition of time-aggregated model statistics $\bbb[y]$, 
\begin{equation}\label{eq:globalIPfin}
 \bbb[y] = \mmm[G]_{T}(\bbb[\theta]; \bbb[v]^{(0)}).
\end{equation}
For sufficiently large $T$, we use the central limit theorem as  in Section \ref{sec:prob} to approximate this relationship as 
\begin{equation}\label{eq:globalIPinf}
\bbb[y] = \mmm[G]_{\infty}(\bbb[\theta]) +  \sigma, \qquad \sigma \sim N\left(0,\Sigma(\bbb[\theta])\right),
\end{equation}
where $\Sigma(\bbb[\theta])$ is the internal variability covariance matrix for parameters $\bbb[\theta]$. To proceed, we must choose a control value $\bbb[\theta]^*$; for example, we take the mean of the prior distribution. Fixing $\bbb[\theta] = \bbb[\theta]^*$, we generate a realization of $\bbb[y]$.

We now solve a set of inverse problems, with the solution of each providing additional information at a design. Specifically, given $\bbb[y]$, we temporarily ``forget" $\bbb[\theta]^*$, and for any design point $k$, we consider
\begin{equation}\label{eq:restrIPinf}
W_k\bbb[y] = W_k\mmm[G]_{\infty}(\bbb[\theta]) +  \sigma_k, \qquad \sigma_k \sim N(0,W_k\Sigma(\bbb[\theta]) W_k^T),
\end{equation}
where $W_k$ restricts the data space to $k$. The posterior distributions of $\bbb[\theta] \mid W_k \bbb[y]$ for all $k$ obtained by solving \eqref{eq:restrIPinf} informs us about the sensitivities of $\mmm[G]_\infty$ with respect to parameters, when only data at different $k$ is available. To simplify the solution of the inverse problem, we approximate the internal variability covariance matrix $\Sigma(\bbb[\theta])$ by a fixed covariance matrix $\Sigma(\bbb[\theta^*])$. This covariance matrix can be obtained by running a collection of control simulations with parameters fixed to (the known) $\bbb[\theta]^*$ but with different initial conditions.

The utility $U$ of a design $W_k$ is a scalar function reflecting the quality of a given design. The design that maximizes the utility function is known as the optimal design. We choose a utility function by measuring information gain (or uncertainty reduction) in $(\bbb[\theta]\mid W_k\bbb[y])$ relative to the prior, in a form of Bayesian optimal design. We use a common choice of utility function that arises in both the Bayesian and non-Bayesian design literature e.g., \cite{ChaVer95,book:FedHac97,Schneider99a,DroMcGPetRya16}, namely, the inverse of the determinant of the information matrix (i.e., the inverse of the posterior covariance matrix),
\begin{equation}\label{eq:Dutil}
U(W_k) = \Bigl(\det \bigl(\mathrm{Cov}(\bbb[\theta]\mid W_k\bbb[y])\bigr)\Bigr)^{-1}.
\end{equation} 
In practice, the posterior covariance matrix is estimated as the empirical covariance matrix of samples drawn from $\bbb[\theta] \mid W_k \bbb[y]$. This utility fulfills the so-called $D$-optimality criterion; unlike trace-based measures (e.g., $A$-optimal utility functions), it is invariant under arbitrary linear transformations of the parameters, for example, when parameters have different dimensional scales. It has been used in investigations of linear and nonlinear design \cite{DroMcGPet13,DroPetRyaTho14,AleGhaGlo16,AleSai18} and particularly in the context of sensor placement \cite{Uci00,FedUci07}. For linear forward maps and Gaussian priors, maximization of this utility is equivalent to maximization of the expected Kullback-Leibler divergence (KLD), a relative entropy measure  \cite{HuaMar13,CooGibGil08,KimLuMyuPitSte14}. While KLD has beneficial mathematical properties, especially for highly non-Gaussian posteriors \cite{Pan05}, it is difficult to evaluate, especially in high-dimensional problems e.g., \cite{HuaMar13}.

\subsection{Synthesis: Targeted uncertainty quantification algorithm}\label{sec:over_alg}

The combined algorithm for targeted uncertainty quantification consists of two stages: first, finding an optimal design point $\tilde{k}$ in a design stage and, second, evaluating parameter uncertainty with data from $\tilde{k}$ in an uncertainty quantification stage. Let $D$ be the finite index set for the set of design points, and define $W_k$ to be the restriction map for any $k\in D$. The two stages then are as follows:
\begin{enumerate}
\item The design stage consists of the following steps:
\begin{enumerate}
\item Generate a sample of GCM simulated data $\bbb[y] = \mmm[G]_T(\bbb[\theta]^*;\bbb[v]^{(0)})$, and estimate the internal variability covariance matrix $\Sigma(\bbb[\theta]^*)$. We approximate $\Sigma(\bbb[\theta])$ as $\Sigma(\bbb[\theta]^*)$.
\item For each $k \in D$, solve \eqref{eq:restrIPinf}, in parallel, for the posterior of $(\bbb[\theta]\mid W_k\bbb[y])$, using the CES-type algorithm described in Appendix \ref{sec:CES}.
\item For each $k \in D$, calculate the utility $ U(W_k)$ from \eqref{eq:Dutil} and choose the optimal design \[\tilde k =  {\arg\max}_{k\in D}U(W_k).\]
\end{enumerate}
\item The uncertainty quantification stage consists of the following steps:
 \begin{enumerate}
 \item At the optimal design point $\tilde k$, obtain a sample $\bbb[z]_{\tilde{k}}$. 
 \item Solve the inverse problem \eqref{eq:localIPinfa}  for the posterior distribution of $(\bbb[\theta] \mid \bbb[z]_{\tilde{k}})$. 
\end{enumerate}
\end{enumerate}

This algorithm could be used as one iteration of a workflow loop where, for example, the posterior distribution $(\bbb[\theta] \mid \bbb[z]_{\tilde{k}})$ can be used to inform a new choice of $\bbb[\theta]^*$.

The complexity of the first stage grows linearly with the candidate design points $k$ because we only consider one point at a time. However, if one wishes to choose a design composed of $K$ simultaneous points  from a set $D$, a combinatorial problem arises, with complexity growing like $|D|!/((|D|-|K|)!|K|!)$---a common problem in the related field of sensor placement design \cite{BraVan01,FedUci07}. This will become prohibitively costly to solve by brute force, even in parallel. We focus on the algorithm for single design points $k$ for now, addressing scaling questions in the discussion section. 
 
\section{Idealized GCM and experimental setup} \label{sec:GCM}

\subsection{Idealized GCM, parameters, and priors}

To demonstrate the algorithm in a simplified setting, we use the idealized aquaplanet GCM described by \cite{Frierson06a} and \cite{Frierson07b} with the modifications introduced by \cite{OGorman08b}. The idealized GCM uses the spectral transform dynamical core of the Flexible Modeling System, developed at the Geophysical Fluid Dynamics Laboratory. We use a coarse spectral resolution of T21 (32 latitude points and 64 longitude points on the Gaussian tranform grid). The vertical is discretized with finite differences with 20 equally spaced sigma levels \cite{Simmons81}. The time discretization uses a second-order leapfrog method with a Robert-Asselin-Williams filter \cite{Williams11a}. The GCM's atmosphere is coupled to a 1-m thick slab ocean, and it uses a  two-stream gray radiation scheme. Convection is represented by a simple quasi-equilibrium moist convection scheme, which relaxes temperature and specific humidity toward moist-adiabatic reference profiles with a fixed relative humidity RH \cite{Frierson07b}. The timescale with which the temperature and specific humidity relax to their respective reference profiles is given by the parameter $\tau$. The parameters RH and $\tau$ are the focus of this study. 

Since the GCM has no topography or other asymmetries at the surface, its statistics are zonally symmetric. With fixed insolation at the top of the atmosphere, the statistics are also statistically stationary. Prescribing seasonally (but not diurnally) varying insolation generates seasonally varying (cyclostationary) statistics, with symmetry between the northern and southern hemisphere (i.e., winter in the northern hemisphere winter is statistically identical to winter in the southern hemisphere) \cite{bordoni08a, DunHowSch22}. \cite{Dunbar21a} and \cite{DunHowSch22} have shown that the parameters  RH and $\tau$ of the convection parameterization in the GCM can be calibrated in the stationary and cyclostationary regimes. Here we want to determine optimal designs for learning about these parameters in the two regimes. 

The priors for these parameters are taken to be logit-normal and lognormal distributions, $\mathrm{RH}\sim \mathrm{Logitnormal}(0, 1)$ and $\tau\sim \mathrm{Lognormal}(12~\mathrm{h},  (12~\mathrm{h})^2)$. That is, we define the invertible transformation
\[
\mathcal{T}(\mathrm{RH}, \tau) = \left(\mathrm{logit}(\mathrm{RH}),~\ln\left(\frac{\tau}{1~\mathrm{s}}\right)\right),
\]
which transforms each parameter to values along the real axis. We label the transformed (or computational) parameters as $\bbb[\theta]= \mathcal{T}(\mathrm{RH}, \tau)$. The untransformed (or physical) parameters (relative humidity and timescale) are uniquely defined by $\mmm[T]^{-1}(\bbb[\theta])$. 
We apply our calibration methods in the space of the transformed parameters $\bbb[\theta]$, where priors are unit-free, normally distributed, and unbounded; meanwhile, the idealized GCM uses the physical parameters $\mmm[T]^{-1}(\bbb[\theta])$, with $\mathrm{RH}\in[0,1]$ and $\tau \in [0,\infty)$. In this way, the prior distributions enforce physical constraints on the parameters. 

\subsection{Objective function for parameter learning}\label{sec:obj_func}

We learn from statistics of model output that are known to be sensitive to the convection parameters. We have knowledge about these sensitivities from a body of previous studies that used this idealized GCM e.g., \cite{OGorman08b, OGorman08c, OGorman09a, bordoni08a, Schneider10a, Merlis11a, OGorman11a, Kaspi11a, Kaspi13c, Levine15a, Bischoff14a, Wills17a, Wei18a}. We know, for example, that the convection scheme primarily affects the atmospheric thermal stratification in the tropics, with weaker effects in the extratropics \cite{Schneider08c}. We also know that the relative humidity parameter RH in the convection scheme controls the humidity of the tropical free troposphere but has a weaker effect on the humidity of the extratropical free troposphere \cite{OGorman11b}. Thus, we expect tropical circulation statistics to be especially informative about the parameters in the convection scheme. However, convection plays a central role in intense precipitation events at all latitudes \cite{OGorman09a,OGorman09b}, so we expect statistics of precipitation intensity to be informative about convective parameters, and in particular to contain information about the relaxation timescale $\tau$.

As statistics to learn from, we choose averages of the free-tropospheric relative humidity, of the precipitation rate, and of a measure of the frequency of intense precipitation. We use averages over $T=90~\mathrm{days}$ in both the statistically stationary and seasonal cycle simulations. We exploit the statistical zonal symmetry in the GCM by taking zonal averages in addition to the time averages. The relative humidity data are evaluated at $\sigma=0.5$ (where $\sigma = p /p_s$ is pressure $p$ normalized by the local surface pressure $p_s$), the precipitation rate is taken daily, and as a measure of the frequency of intense precipitation, we use the frequency with which daily precipitation exceeds the latitude-dependent 90th percentile of precipitation rates in a long (18000 days) control simulation. We hence have 3 statistics, each a function of the 32 latitude points on the spectral transform grid, resulting in a 96-dimensional output vector $\mmm[H]_{T}$. In the statistically stationary case, we take the forward map $\mmm[G]_T = \mmm[H]_T$.

For the simulations with a seasonal cycle, $\mmm[H]_T$ is not statistically stationary but is cyclostationary over multiples of a year. The year length in the GCM is $360~\mathrm{days}$. We stack four 90-day seasons of data together \cite{DunHowSch22} and define the forward map
\[
\mmm[G]_{T}(\bbb[\theta];\bbb[v]^{(0)}) = [\mmm[H]_{T}(\bbb[\theta];\bbb[v]^{(0)}), \dots, \mmm[H]_{T}(\bbb[\theta];\bbb[v]^{(3)}) ]
\]
over a one-year cycle (360 days), where $\bbb[v]^{(i)}$ is the model state at the beginning of each 90-day long season labelled $i=0,1,2,3$. With this batching, we have now constructed stationary statistics for the stacked data. The theory of Section~\ref{sec:method} applies, and our inverse problems can be formulated in the seasonally varying case.

\subsection{Design choices} \label{sec:des_choice}

In the stationary GCM setting, we aggregate statistics temporally and zonally. Thus, a local design implies a restriction to certain latitudes. Recall our discretization has 32 discrete latitudes. We therefore choose a design space that contains sets of $\ell$ consecutive discrete latitudes, indexed from south to north poles. In the stationary experiments, we focus on the case $\ell = 1$. 

In the seasonally varying setting, we still aggregate temporally and zonally, but we also stack the seasons in a vector. We define a local design by indexing both a restriction to a season and a restriction to certain latitudes. We choose a design space that contains sets of $\ell$ consecutive discrete latitudes, collected season by season, indexed from south to north poles. In the seasonal experiments, we  focus on the case $\ell = 1$. 

For additional design scenarios in the stationary setting, we consider cases with wider design stencils, $\ell=3$, in Appendix \ref{sec:3stencil}, and we consider cases with shorter averaging periods, $T=30$ days, in Appendix \ref{sec:30day}.

\subsection{Synthetic data and noise}\label{sec:synth_data}

We generate limited-area data $\bbb[z]_k$ with the idealized GCM itself at a fixed parameter vector $\bbb[\theta]^\dagger$, adding Gaussian noise $\delta_k$ with zero mean and covariance matrix $\Delta$ as in \eqref{eq:localIPinfa}. One interpretation of this added noise is that it plays the role of an artificial corruption of $\mmm[S]_T(\bbb[\theta]^\dagger;k)$, with unbiased model error $\delta_k$ that plays the same role as additional observational noise \cite{Kennedy01a}. One can obtain unbiased $\delta_k$ by inclusion of models for structural model error within $\mmm[S]_T$, for example, learned error models that enforce conservation laws and sparsity \cite{LevStu21-pre,Schneider22v}. The inverse problem \eqref{eq:localIPinfa} can be written as 
\begin{equation}\label{eq:localIPinf}
 \bbb[z]_k = \mmm[S]_{\infty}(\bbb[\theta];k) + \gamma_k,\qquad \gamma_k \sim N(0,W_k(\Sigma(\bbb[\theta]) +\Delta)W_k^T).
\end{equation}
We construct the measurement error covariance matrix $\Delta$ to be diagonal with entries $d_i^2 = \Delta_{ii}>0$, where $i$ indexes over data type (three observed
quantities) and over the discrete latitudes, 
\begin{equation}\label{eq:gammadef}
\Sigma+\mathop{\mathrm{diag}(d_{i}^2)} = \Sigma + \Delta.
\end{equation}
We choose $d_i$ so that it is proportional to the mean $\mu_i$ of the variable in question, with a proportionality factor $C_{\max}=0.1$. To prevent the noise from becoming so large that the variables can cross a physical boundary $\partial \Omega_i$ (e.g., relative humidity becoming negative), we limit the noise standard deviation to a factor $C=0.2$ times the distance between the approximate 95\% noise confidence interval and the physical boundary:
\[
d_i = \min\Big(C \min\left(\text{dist}(\mu_i + 2\sqrt{\Sigma_{ii}}, \partial \Omega_i),\text{dist}(\mu_i - 2\sqrt{\Sigma_{ii}}, \partial \Omega_i)\right), C_{\max}\mu_i \Big).
\]  

In our proof-of-concept experiments, we generate a sample of ground truth data, $\bbb[z]_k$, and its variability, by carrying out a set of control simulations, with the parameters fixed to values $\bbb[\theta]^\dagger$, where $\mmm[T]^{-1}(\bbb[\theta]^\dagger) =  (0.7,2~\mathrm{h})$ are standard values used in previous studies \cite{OGorman08b}. We use this set of control simulations to estimate the restricted covariance matrix $W_k\Sigma(\bbb[\theta])W_k^T \approx W_k\Sigma(\bbb[\theta]^\dagger)W_k^T$ for any $k$.  In the statistically stationary case, we carry out control simulations for 200 windows of length $T=90~\mathrm{days}$, after discarding the first 50 months for spin-up, and we calculate the sample covariance matrix $\Sigma(\bbb[\theta^\dagger])$ from the  200 samples. Here, $W_k\Sigma(\bbb[\theta]^\dagger)W_k^T$ is a symmetric matrix whose size depends on the design space; it represents noise from internal variability in the 90-day time averages. In the seasonally varying case, we carry out a control simulation for 150 years, discarding the first 4 years for spin-up, and obtain the sample covariance matrix $\Sigma(\bbb[\theta^\dagger])$ from the stacked seasonal ($T=90~\mathrm{days}$) averages.  In the seasonal case, $W_k\Sigma(\bbb[\theta]^\dagger)W_k^T$ is a symmetric matrix whose size depends on $4$ times the design space. We add a small regularization term of $10^{-4}$ to the diagonal of $\Sigma(\bbb[\theta]^\dagger)$ to prevent zero variability, which occurs due to finite-time averages of intense precipitation. In practical implementations of this method, good estimates of the local variability that we represent by $W_k\Sigma(\bbb[\theta]^\dagger)W_k^T$ can be obtained from the observed climatology of the statistics of interest, instead of estimating them from a control simulation of the GCM. 

In the data acquisition algorithm, we require a sample of data $W_k\bbb[y]$, and its variability, for different $k$. To obtain this, we use a set of control simulations of the GCM in which we fix the parameters to the prior mean $\bbb[\theta]^*$, the value used to generate $\bbb[y]$, equivalent to the physical values $\mmm[T]^{-1}(\bbb[\theta]^*) =  (0.5,7~\mathrm{h})$. In the stationary case, the three latitude-dependent fields evaluated at 32 latitude points produce a $96 \times 96$ symmetric matrix $\Sigma(\bbb[\theta]^*)$, representing noise from internal variability in 90-day averages.
Similarly, in the seasonal case, the stacked statistics produce a $384 \times 384$ symmetric matrix $\Sigma(\bbb[\theta]^*)$, and since $T=90~\mathrm{d}$, $\Sigma(\bbb[\theta]^*)$ represents noise from internal variability in 90-day averages. We again add a small regularization term of $10^{-4}$ to the diagonal of $\Sigma(\bbb[\theta]^*)$. In both cases, we estimate $\Sigma(\bbb[\theta]) \approx \Sigma(\bbb[\theta]^*)$ in the optimal design stage of the algorithm.

\begin{figure}[h]
 \centering
\includegraphics[width=\textwidth]{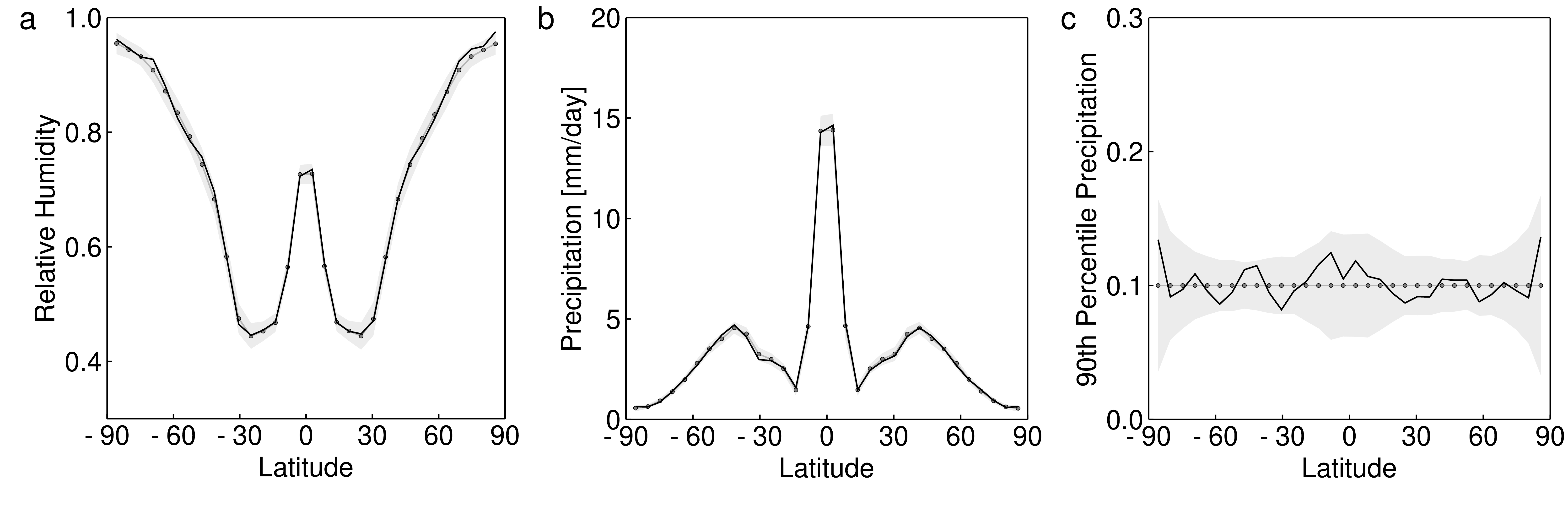}
\caption{Aggregated climate statistics  in the statistically stationary control simulation, with parameters set to the mean of the prior $\bbb[\theta]^*$. The mean (grey lines) and 95\% confidence intervals (shading) of the data are plotted against latitude. 
One realization of the 90-day averaged data is shown (black line). No noise is added here.}
\label{f:data_meanparam}
\end{figure}

\begin{figure}[h]
 \centering
\includegraphics[width=\textwidth]{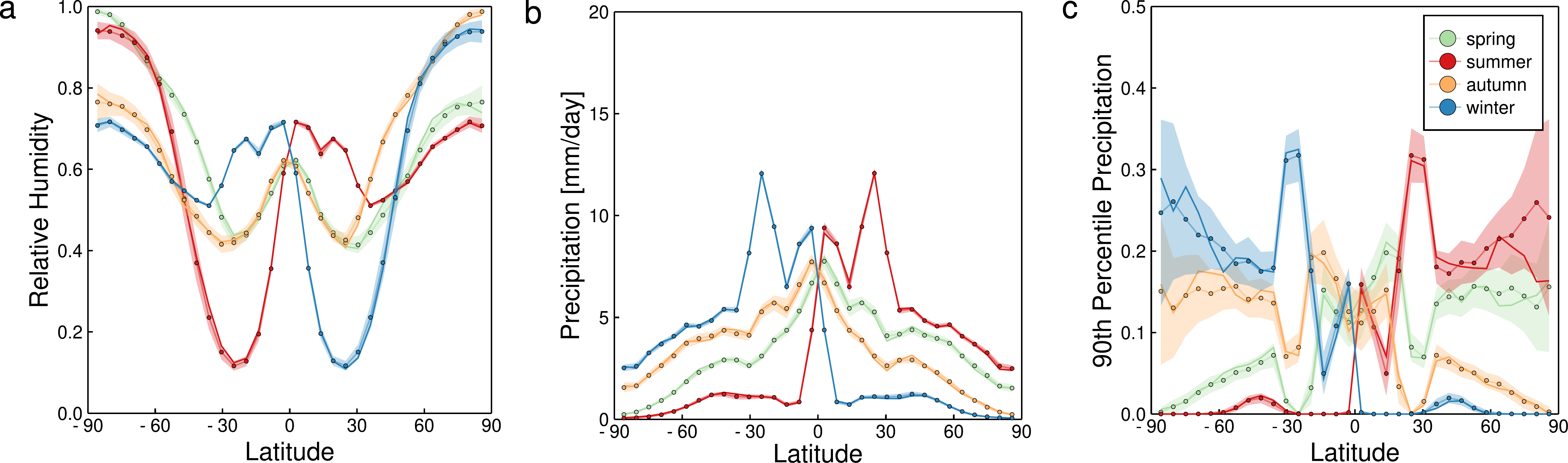}
\caption{Aggregated climate statistics in the seasonally varying control simulation, with parameters set to the mean of the prior $\bbb[\theta]^*$. The mean (solid lines) and 95\% confidence intervals (shading) of the data are plotted against latitude, with the different colors for different seasons, with the labels referring to the northern hemisphere. The infinite-time statistics between the two hemispheres are identical, so differences between, e.g., northern and southern hemisphere winter or summer are indicative of sampling variability from finite-time averages. No noise is added here.}  
\label{f:seasonal_data_meanparam}
\end{figure}

\begin{figure}[h]
 \centering
\includegraphics[width=\textwidth]{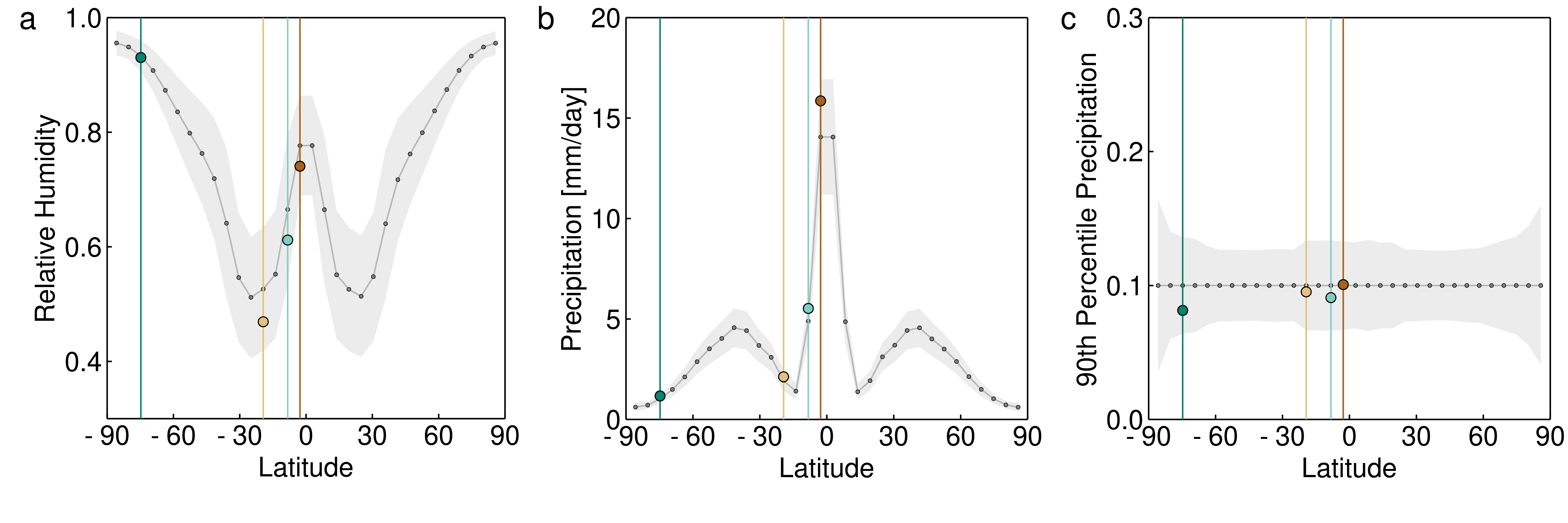}
\caption{Aggregated climate statistics in the statistically stationary control simulation using the ground truth parameters $\bbb[\theta]^\dagger$. The mean (grey lines) and 95\% confidence intervals (shading) of the data are plotted against latitude. Noise mimicking observational and/or model error is added. Each colored disc represents a 90-day realization of GCM data coming from a different design (latitude) used in the experiment.}
\label{f:data_phys}
\end{figure}

\begin{figure}[h]
 \centering
\includegraphics[width=\textwidth]{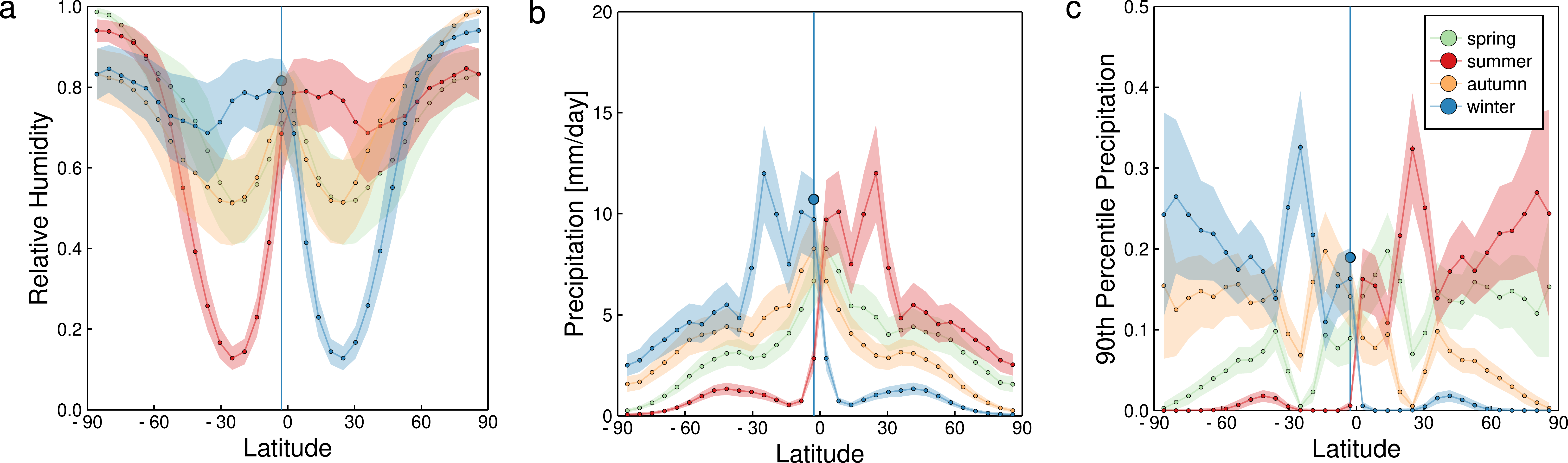}
\caption{Aggregated climate statistics in the seasonally varying control simulation using the ground truth parameters $\bbb[\theta]^\dagger$. We added noise mimicking observational and/or model error. The mean (solid lines) and 95\% confidence intervals (shading) of the data are plotted against latitude, with the colors indicating different seasons, referenced to the northern hemisphere. The blue vertical line indicates the location and season (northern winter) in which we observe the data for uncertainty quantification; the specific 90-day realization of GCM data for the one-latitude design is given by the blue disc.}
\label{f:seasonal_data_phys}
\end{figure}

The mean and 95\% confidence interval of the data at $\bbb[\theta]^*$, with covariance constructed from $\Sigma(\bbb[\theta]^*)$, are shown in Figure~\ref{f:data_meanparam} for the statistically stationary case and in Figure~\ref{f:seasonal_data_meanparam} for the seasonally varying case. The black (stationary) and colored (seasonal) solid lines illustrate a realization of the data for one initial condition. Similarly, the mean and 95\% confidence interval of the data at $\bbb[\theta]^\dagger$, with noise added with covariance matrix $\Delta + \Sigma(\bbb[\theta]^\dagger)$, are shown in Figure~\ref{f:data_phys} for the stationary and in Figure~\ref{f:seasonal_data_phys} for the seasonally varying case.

\section{Results}\label{s:results}

\subsection{Stationary statistics}\label{ss:stationary}

\begin{figure}[h]
 \centering
\includegraphics[width=0.65\textwidth]{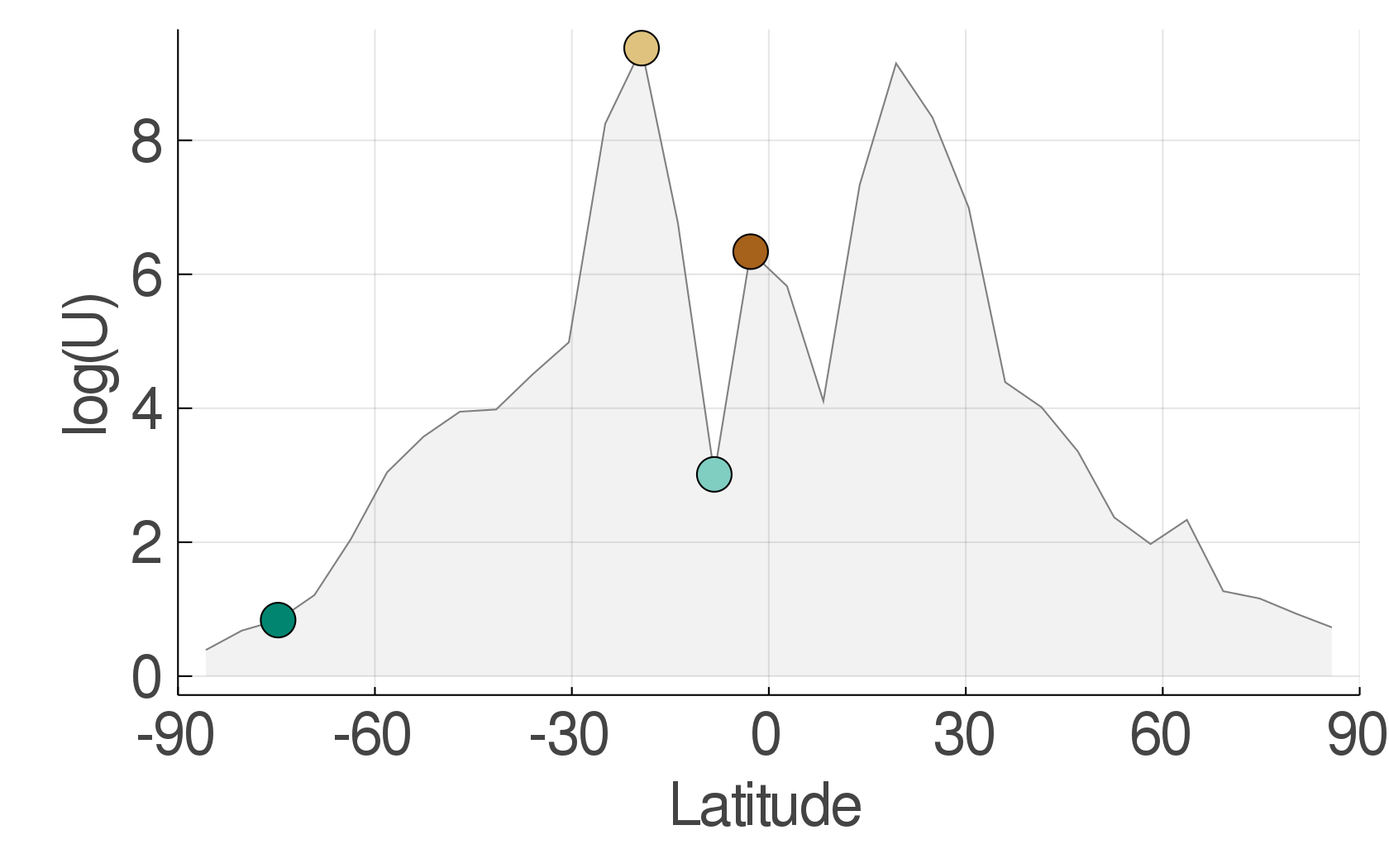}
\caption{Logarithm of the data utility as a function of latitude, with designs corresponding to a single latitude. The colored discs signify the four representative designs indicated in Fig.~\ref{f:data_phys}, which are used in the uncertainty quantification experiment.}
\label{f:util_1stencil}
\end{figure}

\begin{figure}[h]
    \centering
    \includegraphics[width=0.9\textwidth]{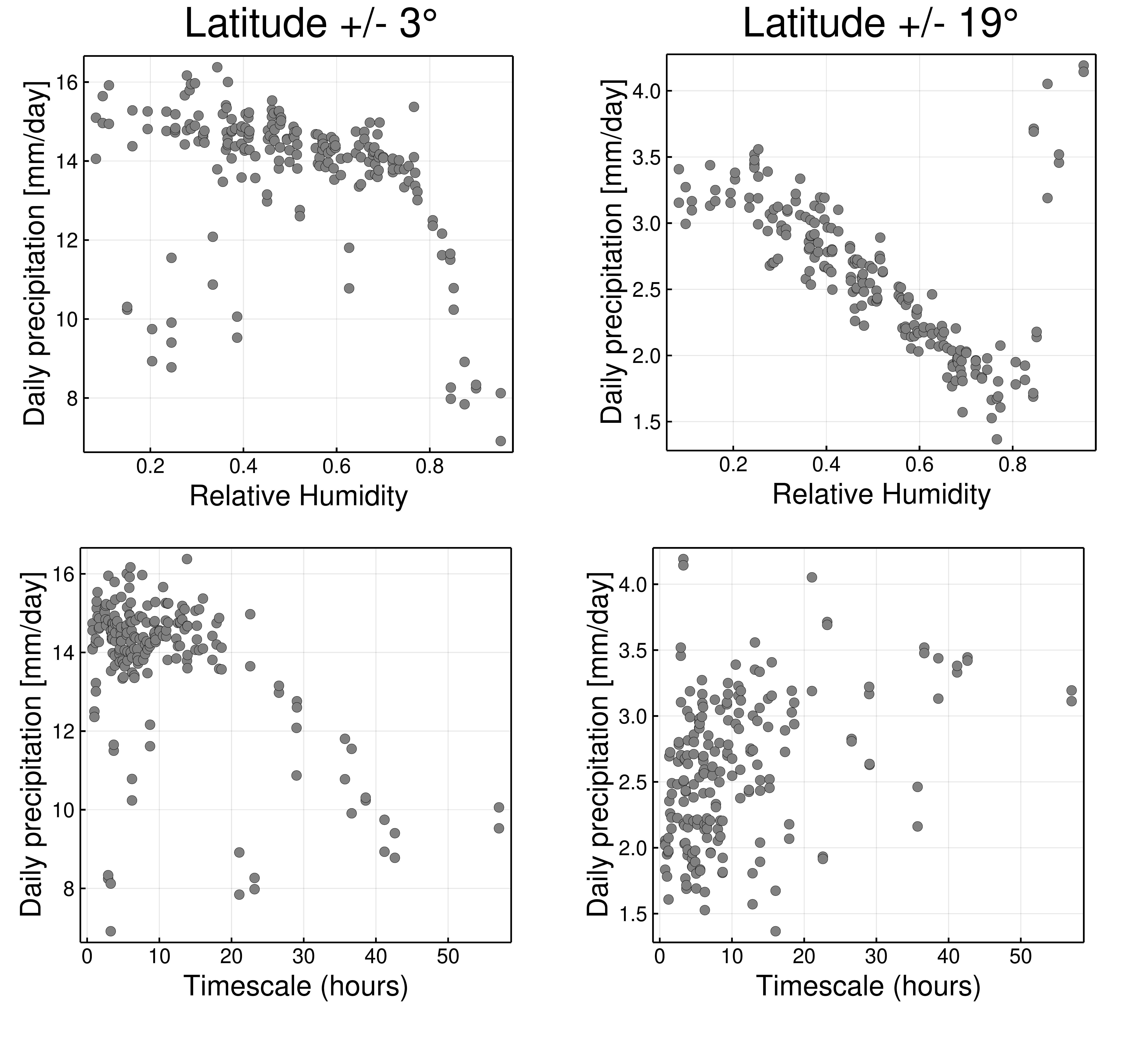}
    \caption{Daily precipitation rates at equatorial (left) and subtropical (right) latitudes, plotted against the relative humidity and relaxation timescale in the convection scheme. The scatter plots are generated by sampling independently from the prior distribution for the two parameters and then projecting into each parameter dimension. }
\label{f:precip_scatter}
\end{figure}

We first apply the optimal design algorithm to the statistically stationary GCM. The logarithm of the utility function is shown in Figure~\ref{f:util_1stencil}. The extent to which hemispheric symmetry of the statistics is broken in Figure~\ref{f:util_1stencil} is an indication of sampling variability, as the infinite-time GCM statistics are hemispherically symmetric. The design landscape appears surprising, as precipitation and parameterized tendencies from convection are largest in the ITCZ (within $\pm3^\circ$ of the equator), and one may expect the optimal region to be in the ITCZ as well. Our algorithm indicates that the equatorial region is indeed a good location, but larger utilities are found at latitude $\pm 19^\circ$, near the precipitation minima under the descending branches of the Hadley circulation in this model. Indeed, daily precipitation rates at this subtropical latitude correlate more strongly with the relative humidity parameter in the convection scheme than in the equatorial latitudes (Figure~\ref{f:precip_scatter}). With designs focused on a single latitude ($\ell=1$), this region is indicated to be most informative. With wider design stencils ($\ell=3$), the algorithm's aligns closer with intuition, placing optimal utility near the equator (Figure~\ref{f:util_3stencil}).

\begin{figure}[h]
    \centering
    \includegraphics[width=0.8\textwidth]{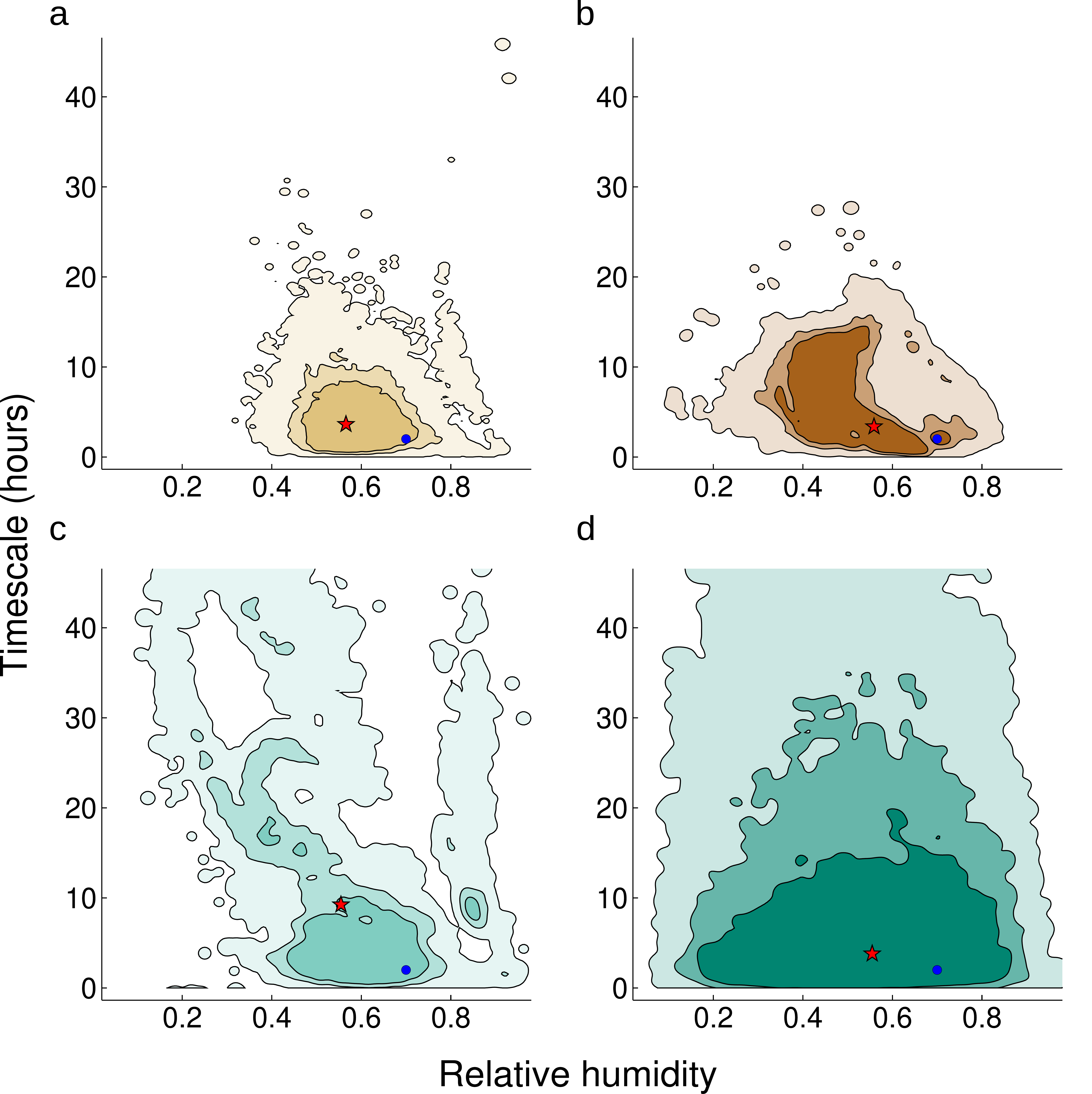}
     \caption{Posterior distributions for convection parameters learned from data restricted to different design points. The drawn contours bound 50\%, 75\%, and 99\% of the distribution. Panels a--d correspond to designs at latitudes $-19^\circ$, $-3^\circ$, $3^\circ$, and $-75^\circ$, ordered according to decreasing utility in Figure~\ref{f:util_1stencil}. The true utility of these distributions are $26.4$, $13.9$, $4.4$, and $1.7$. The true parameter values in the control simulation are given by the blue circle. The parameters found to be optimal in the calibration scheme (given a single random realization of data) are given by the red star in each case.}
    \label{f:posterior_phys}
\end{figure}

We validate our optimal choice by solving \eqref{eq:localIPinf} at four representative design choices, at latitudes $-19^\circ$, $-3^\circ$, $-8^\circ$, and $-75^\circ$, (in decreasing order of utility) shown as colored discs in Figure ~\ref{f:util_1stencil}. The samples of climate statistics used at each latitude are shown in Figure~\ref{f:data_phys} (colored discs). Density plots of the posterior distributions at each latitude are shown in Figure~\ref{f:posterior_phys}. Each panel shows the density contours bounding 50\%, 75\%, and 99\% of the posterior distribution, shaded dark to light; the priors are largely uninformative and have been excluded from the plots. The panels a---d are ordered by decreasing utility from Figure~\ref{f:util_1stencil}, which is a predictor of information content based on uncertainty at the prior mean $\bbb[\theta]^*$. The true utilities of the posterior distributions $\bbb[\theta]^\dagger \mid \bbb[z]_k$ are $26.4$, $13.9$, $4.4$, and $1.7$. Thus, the order of predicted information content reflects the order of actual information content. Visually, we see an increased area covered by the different contours for less informative distributions. However, the prediction of the ordering of utilities does not extend to providing accurate prediction of the actual utility value, due to the additional error inflation present in the true data and sampling error. Physical intuition, positing the equatorial region as the optimal target location, would lead to a reasonable design with a utility of 13.9 (Figure~\ref{f:posterior_phys}b), around half that of the optimal design (26.4). A poor guess, positing high latitudes as the optimal target location, would lead to only moderate improvements relative to the prior (Figure~\ref{f:posterior_phys}d),  with a utility of 1.7 that is around a factor 20 smaller than that for the optimal design. With wider design stencils, the optimally informative location is predicted to be closer to the equator (Figure~\ref{f:util_3stencil}) As observed in other investigations \cite{Dunbar21a}, the posterior distributions are subject to variability due to the finite-time sampling and the inflation. However, all distributions capture the true parameter values within $50\%$ of the posterior mass.

When the climate statistics that are used are based on shorter-term averages and hence are noisy, the targeting algorithm, as expected, can become less effective, and parameter posteriors can become more multimodal (Figure~\ref{f:30day_posterior_stencils}).

\subsection{Seasonally varying statistics} \label{sec:seas_GCM}

\begin{figure}[h]
 \centering
\includegraphics[width=0.65\textwidth]{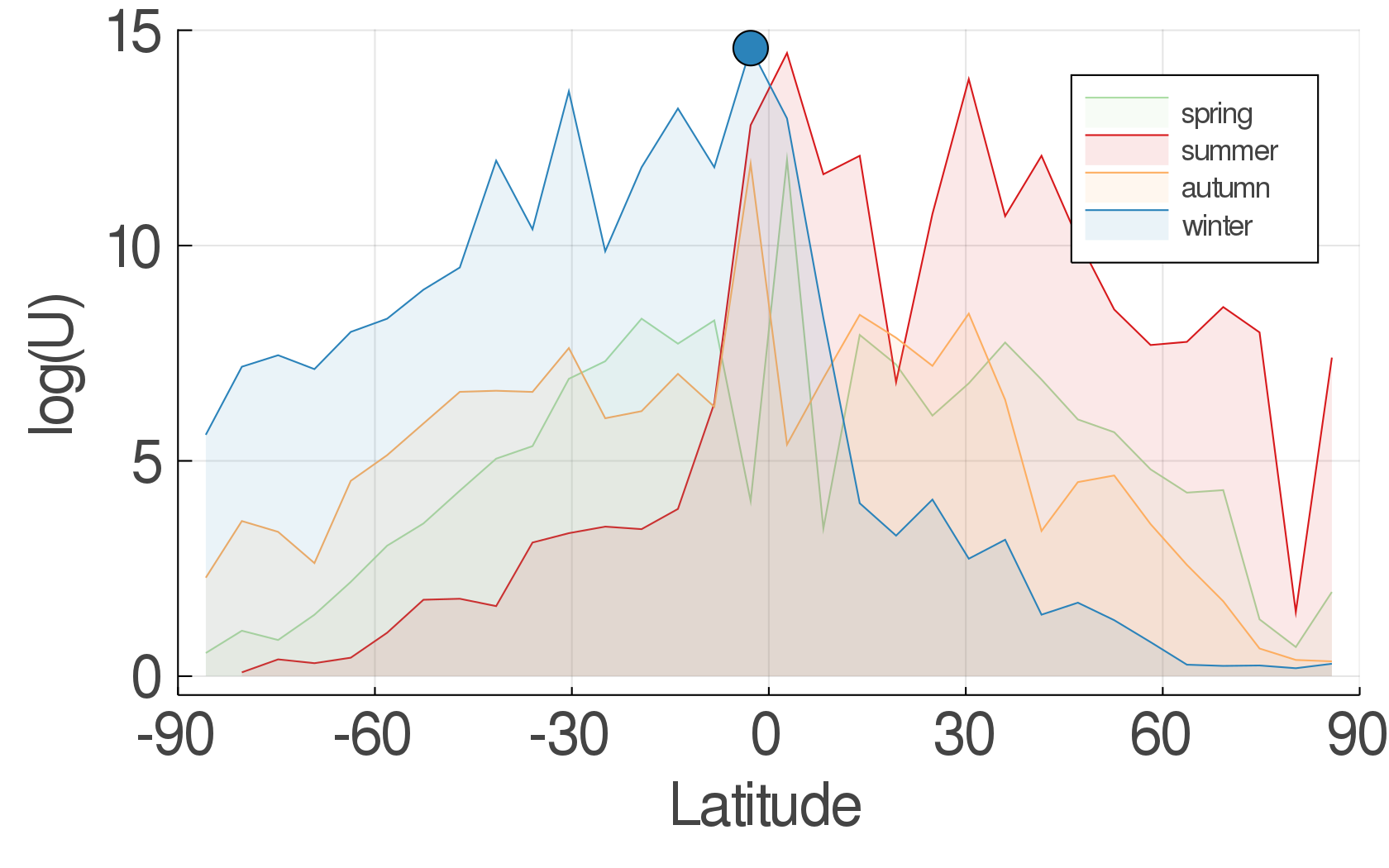}
\caption{Logarithm of the data utility plotted against latitude (1 design per latitude). The shading represents the (northern) season over which data was averaged. The blue disc signifies that an equatorial latitude in northern winter maximizes the utility function across all locations and seasons.}
\label{f:seasonal_utilities}
\end{figure}

In the seasonally varying case, we choose the optimal design with the algorithm in Section~\ref{sec:over_alg} applied to the data stacked in seasons. Figure~\ref{f:seasonal_utilities} shows the logarithm of the utility function. Hemispheric and seasonal asymmetries are evident here. In northern winter, latitudes just south of the the equator ($-3^\circ$) optimize the design, in the vicinity of the ITCZ. Conversely, in northern summer, latitudes just north of the equator ($3^\circ$) optimize the design, again in the vicinity of the seasonally migrating ITCZ. Additional peaks in the data utility can be seen around $30^\circ$, in the summer subtropics and again near the descending branch of the Hadley circulation. The equinox seasons have less utility at the optimal designs ($3^\circ$ and $-3^\circ$). Because the equinoctial Hadley cells and ascent regions in the ITCZ are less pronounced than the solstitial Hadley cells \cite{Schneider10a}, utility is more spread out across the latitudes.

\begin{figure}[h]
 \centering
\includegraphics[width=0.8\textwidth]{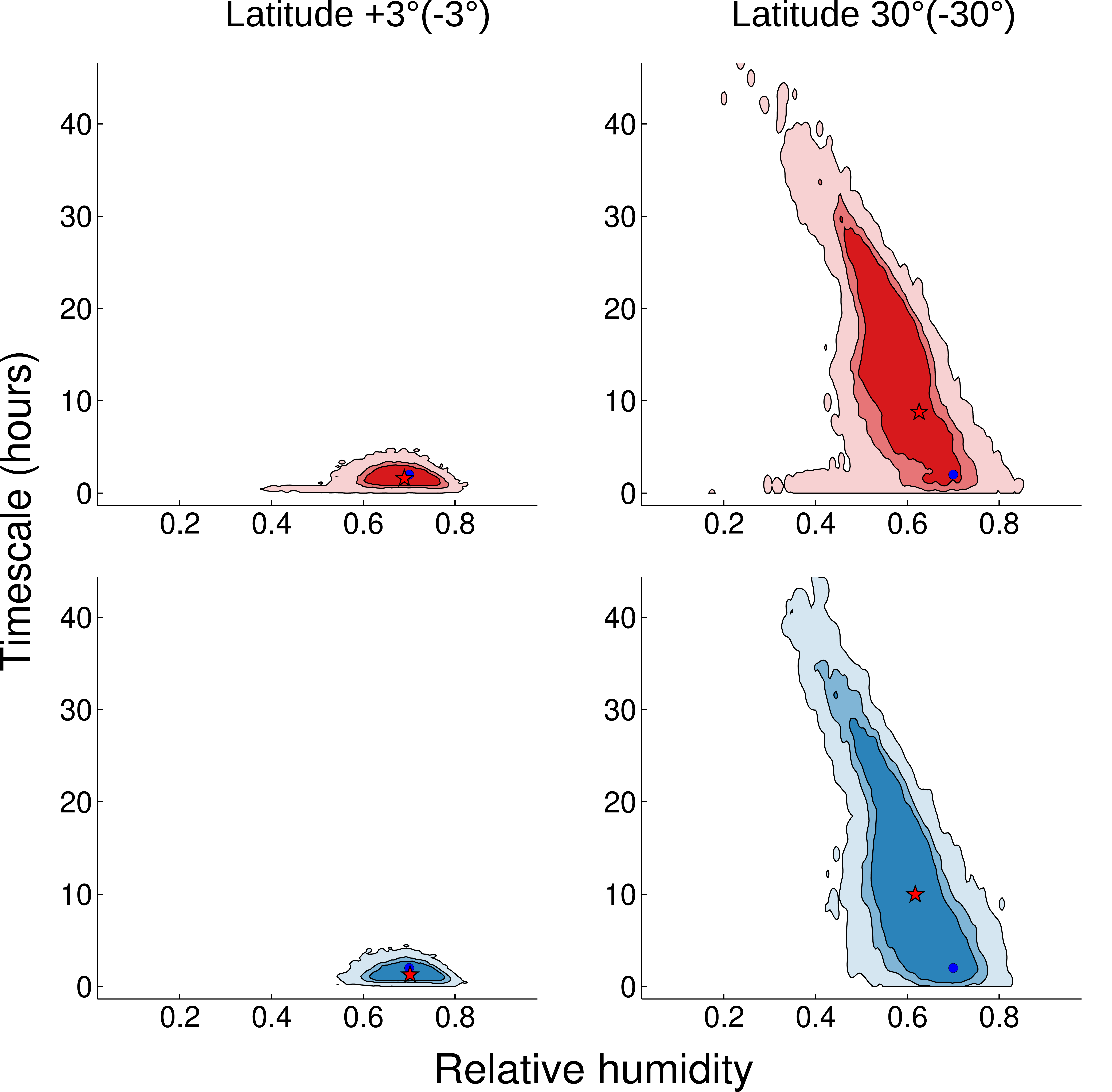}
\caption{Posterior distribution obtained from using data at the optimal latitudes ($\pm3^\circ$, left) and second-optimal latitudes ($\pm30^\circ$, right). The top row corresponds to data targeted to northern summer in the northern hemisphere, and the bottom row corresponds to data targeted to southern summer in the southern hemisphere. Contours bound 50\%, 75\%, and 99\% of the distribution (in decreasing color saturation). The true utility of the northern summer distributions are (left: 131.9, right: 47.9), and southern summer distributions are (left: 154.7, right: 39.5). The true parameter values in the control simulation are given by the blue circle. The parameters found to be optimal in the calibration scheme (given a single random realization of data) are given by the red star in each case.}
\label{f:seasonal_posterior_phys}
\end{figure}

\begin{figure}[h]
 \centering
\includegraphics[width=0.8\textwidth]{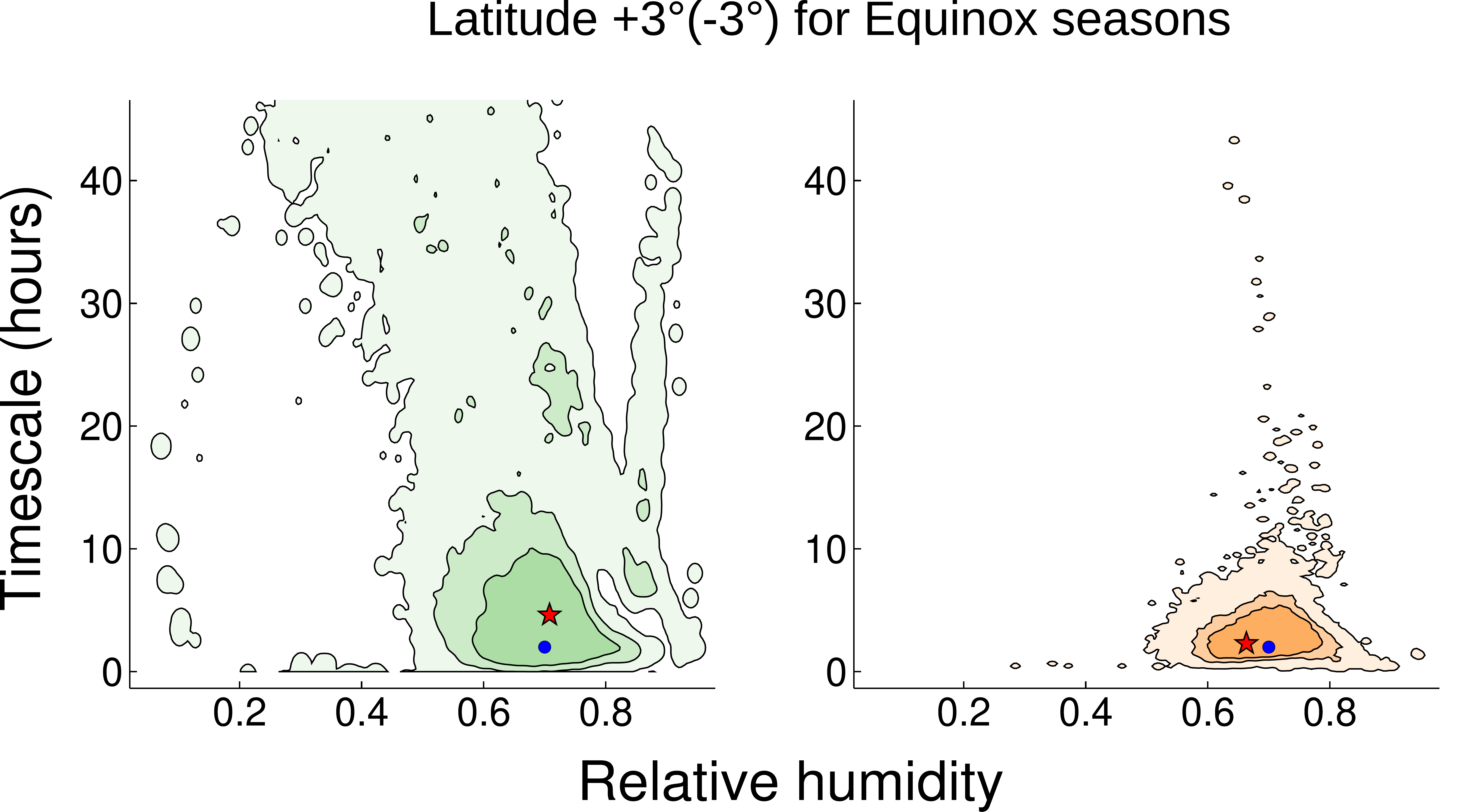}
\caption{Posterior distribution obtained from uncertainty quantification using data targeted at the optimal latitude ($\pm3^\circ$) from each equinox season. Contours bound 50\%, 75\%, and 99\% of the distribution (in decreasing color saturation). The northen spring (at latitude $+3^\circ$) distribution has utility 54.8, while northern autumn (at latitude $-3^\circ$) has utility 89.7. The true parameter values in the control simulation are given by the blue circle. The parameters found to be optimal in the calibration scheme (given a single random realization of data) are given by the red star in each case.}
\label{f:seasonal_posterior_phys_equinox}
\end{figure}

We solve the analogue inverse problem \eqref{eq:localIPinf} as in the statistically stationary case with a sample of data taken at latitudes of $\pm3^\circ$ or $\pm30^\circ$, corresponding to the first and second peaks of utility for the solstice seasons. The posterior distributions are collected in Figure~\ref{f:seasonal_posterior_phys}, colored by season.  In general, the true parameter values lie within 50\% of the posterior mass in each case. The utilities at the optimal latitudes in northern summer and winter are 131.9 and 154.7, respectively. In contrast, the utilities corresponding to the secondary peaks in the subtropics are 47.9 and 39.5 for northern summer and winter, respectively. As in the statistically stationary case, the design with highest predicted utility (northern winter at $3^\circ$) indeed has highest utility. Visually we see symmetry between these seasons, with qualitatively similar distributions in the opposing hemispheres for northern summer and winter. For the equinox seasons, from data sampled at their respective optimal latitudes of $+3^\circ$ and $-3^\circ$ (Figure~\ref{f:seasonal_posterior_phys_equinox}), we see lower utilities of 89.7 and 54.8 for northern fall and spring, respectively, and we see asymmetry most likely indicating sampling variability, because the infinite-time GCM statistics are hemispherically symmetric. In this seasonally varying setting, we again observe that our targeted data acquisition algorithm is a good predictor of informativeness of additional data for learning about the convection parameters. 

\section{Conclusions and Discussion}\label{sec:conclusion}

We have presented a novel framework for automated optimal data acquisition to calibrate a global model. The framework can be used with computationally expensive and chaotic (noisy) GCMs, whose derivatives may not be available. The data are assumed to be accessible only at limited locations and at different times of year. Given a global simulation, we use parameter uncertainty information provided by the CES algorithm to guide our choice of design (when and where we target data acquisition). We have demonstrated the efficacy of the algorithm for finding optimally informative locations in perfect-model settings in which we generated data with an idealized GCM and learnt about parameters in a convection parameterization. Using statistically stationary or seasonally varying statistics, we have explored both spatial and spatio-temporal designs. 

With the idealized GCM, we have targeted a location and time period at which additional data will produce parameter estimates that minimize uncertainty. In our proof-of-concept with narrow designs consisting of data measured only at a single latitude ($\ell=1$), the automatically targeted optimal location for new data acquisition was, in the seasonal case, in the vicinity of the seasonally migrating ITCZ, with secondary maxima in the summer subtropics. This is consistent with the fact that the convection scheme in the idealized GCM is most important near the ITCZ \cite{OGorman08b}. In the statistically stationary case, regions near the ITCZ are optimal for data acquisition with wider design stencils ($\ell=3$, Appendix \ref{sec:3stencil}). However, in scenarios with narrower designs ($\ell=1$), the subtropical precipitation minimum turns out to be the optimal location, which we confirmed by calibrating convection parameters at this and other locations. We showed that the optimal targeting is limited in its effectiveness when the available data are very noisy (as shown in Appendix \ref{sec:30day} when both the averaging timescale and stencil sizes are reduced). However, the algorithm provides access to the posterior distributions of the parameters, so that this behavior is both diagnosable a posteriori and actionable with successive iterations of the optimal design process (for example, using the current posterior as the prior for a subsequent iteration with additional data). We also showed that although the algorithm correctly predicts the ordering of information content of different sites in many scenarios, it does not necessarily provide an accurate estimate of the actual information content at the sites, due to sampling variability and the additional model error inflation.  

Our algorithm couples the optimization over the design space to the specific application through the posterior distribution of parameters. Therefore, it captures different applications of targeted data acquisition by modifying only the forward map and data entering the loss function to learn this parameter distribution, without changing the algorithm structure. Our framework is thus immediately applicable to the motivating example of automatically targeting embedded high-resolution simulations such as those in  \cite{Shen20a} and \cite{JarSchSheSriZhi21-pre} to regions that are maximally informative about parameterizations. One could even consider targeting observational data acquisition, such as informing choices for new field campaigns e.g., \cite{Ste-etal03,Rauber07a}, or new in-situ observatory locations e.g., \cite{SchSto94}. However, many additional practical considerations beyond the scope of optimal experimental design also play a role in site selection in such cases.

The current algorithm relies on evaluating utilities naively at all design points. Thus, for moderately sized design spaces, the computational cost is dominated by the cost of running the GCM. In practice, if we want to determine $\mathcal{O}(10^3)$ limited-area data acquisition sites optimally within $10^6$ or more possible locations, such naive approaches are inefficient. Instead, one can use more sophisticated optimization algorithms. For determinant based (i.e., $D$-optimal) utilities, this typically requires accelerating the determinant evaluation (and its gradients). Various methods have been developed to do so, e.g., using Laplace approximations \cite{LonScaTemWan13,BecDiaEspLonTem18,ChoMarRue09}, polynomial chaos surrogates \cite{HuaMar14}, optimization of criteria bounds \cite{GahHajTsi17},  fast random determinant approximation \cite{AleGhaPetSta14,AleSai18}, and Gaussian process surrogates \cite{BuaGinKri20,EidKarPag20}. The latter, kernel-based approaches are particularly amenable to our setting, as they give sparse representations of the utility function that are independent of the underlying computational grid. They may offer a way forward in the climate modeling setting. 

As we have presented it here, the algorithm is directly applicable to comprehensive climate models. It will be interesting to explore to what extent application to comprehensive models yields results such as the ones we have seen in the idealized setting: non-obvious optimal locations for targeting computational or observational data acquisition for reducing uncertainties in convection or other parameterization schemes

\section*{Acknowledgments}
We gratefully acknowledge the generous support of Eric and Wendy Schmidt (by recommendation of Schmidt Futures) and the National Science Foundation (grant AGS-1835860). The simulations were performed on Caltech's High Performance Cluster, which is partially supported by a grant from the Gordon and Betty Moore Foundation. AMS is also supported by the Office of Naval Research (grant N00014-17-1-2079). 

\section*{Data Availability} 
All computer code used in this paper is open source. The code for the idealized GCM, the Julia code for the optimal design algorithm, the plotting tools, and the slurm/bash scripts to run both GCM and design algorithms are available at:

https://doi.org/10.5281/zenodo.6679974

\bibliographystyle{siam}

\begin{thebibliography}{100}

\bibitem{AleGhaGlo16}
{\sc A.~Alexanderian, P.~J. Gloor, and O.~Ghattas}, {\em On {B}ayesian {A}- and
  {D}-optimal experimental designs in infinite dimensions}, {B}ayesian
  Analysis, 11 (2016), pp.~671--695.

\bibitem{AleGhaPetSta14}
{\sc A.~Alexanderian, N.~Petra, G.~Stadler, and O.~Ghattas}, {\em {A}-optimal
  design of experiments for infinite-dimensional {B}ayesian linear inverse
  problems with regularized $\ell_0$-sparsification}, SIAM J. Sci. Comput., 36
  (2014), pp.~A2122--A2148.

\bibitem{AleSai18}
{\sc A.~Alexanderian and A.~K. Saibaba}, {\em Efficient {D}-optimal design of
  experiments for infinite-dimensional {B}ayesian linear inverse problems},
  SIAM J. Sci. Comput, 40 (2018), pp.~A2956--A2985.

\bibitem{BecDiaEspLonTem18}
{\sc J.~Beck, B.~M. Dia, L.~F. Espath, Q.~Long, and R.~Tempone}, {\em Fast
  {B}ayesian experimental design: {L}aplace-based importance sampling for the
  expected information gain}, Comput. Methods Appl. Mech. Eng., 334 (2018),
  pp.~523 -- 553.

\bibitem{Bischoff14a}
{\sc T.~Bischoff and T.~Schneider}, {\em Energetic constraints on the position
  of the {I}ntertropical {C}onvergence {Z}one}, J. Climate, 27 (2014),
  pp.~4937--4951.

\bibitem{bishop1999ensemble}
{\sc C.~H. Bishop and Z.~Toth}, {\em Ensemble transformation and adaptive
  observations}, J. Atmos. Sci., 56 (1999), pp.~1748--1765.

\bibitem{Bony06}
{\sc S.~Bony, R.~Colman, V.~M. Kattsov, R.~P. Allan, C.~S. Bretherton, J.-L.
  Dufresne, A.~Hall, S.~Hallegatte, M.~M. Holland, W.~Ingram, D.~A. Randall,
  B.~J. Soden, G.~Tselioudis, and M.~J. Webb}, {\em How well do we understand
  and evaluate climate change feedback processes?}, J. Climate, 19 (2006),
  pp.~3445--3482.

\bibitem{BonDuf05}
{\sc S.~Bony and J.-L. Dufresne}, {\em Marine boundary layer clouds at the
  heart of tropical cloud feedback uncertainties in climate models}, Geophys.
  Res. Lett., 32 (2005).

\bibitem{bordoni08a}
{\sc S.~Bordoni and T.~Schneider}, {\em Monsoons as eddy-mediated regime
  transitions of the tropical overturning circulation}, Nature Geosci., 1
  (2008), pp.~515--519.

\bibitem{Brient16b}
{\sc F.~Brient and T.~Schneider}, {\em Constraints on climate sensitivity from
  space-based measurements of low-cloud reflection}, J. Climate, 29 (2016),
  pp.~5821--5835.

\bibitem{BuaGinKri20}
{\sc P.~Buathong, D.~Ginsbourger, and T.~Krityakierne}, {\em Kernels over sets
  of finite sets using {RKHS} embeddings, with application to {B}ayesian
  (combinatorial) optimization}, in Int. Conf. Artif. Intell. Stat., PMLR,
  2020, pp.~2731--2741.

\bibitem{Cess90a}
{\sc R.~D. Cess, G.~L. Potter, J.~P. Blanchet, G.~J. Boer, A.~D. Del~Genio,
  M.~D{\'e}qu{\'e}, V.~Dymnikov, V.~Galin, W.~L. Gates, S.~J. Ghan, J.~T.
  Kiehl, A.~A. Lacis, H.~Le~Treut, Z.-X. Li, X.-Z. Liang, B.~J. McAvaney, V.~P.
  Meleshko, J.~F.~B. Mitchell, J.-J. Morcrette, D.~A. Randall, L.~Rikus,
  E.~Roeckner, J.~F. Royer, U.~Schlese, D.~A. Sheinin, A.~Slingo, A.~P.
  Sokolov, K.~E. Taylor, W.~M. Washington, R.~T. Wetherald, I.~Yagai, and M.-H.
  Zhang}, {\em Intercomparison and interpretation of climate feedback processes
  in 19 atmospheric general circulation models}, J. Geophys. Res., 95 (1990),
  pp.~16601--16615.

\bibitem{Ces-etal89}
{\sc R.~D. Cess, G.~L. Potter, J.~P. Blanchet, G.~J. Boer, S.~J. Ghan, J.~T.
  Kiehl, H.~L. Treut, Z.-X. Li, X.-Z. Liang, J.~F.~B. Mitchell, J.-J.
  Morcrette, D.~A. Randall, M.~R. Riches, E.~Roeckner, U.~Schlese, A.~Slingo,
  K.~E. Taylor, W.~M. Washington, R.~T. Wetherald, and I.~Yagai}, {\em
  Interpretation of cloud-climate feedback as produced by 14 atmospheric
  general circulation models}, Science, 245 (1989), pp.~513--516.

\bibitem{ChaVer95}
{\sc K.~Chaloner and I.~Verdinelli}, {\em {B}ayesian experimental design: A
  review}, Stat. Sci., 10 (1995), pp.~273--304.

\bibitem{CheOli12}
{\sc Y.~Chen and D.~S. Oliver}, {\em Ensemble randomized maximum likelihood
  method as an iterative ensemble smoother}, Math. Geosci., 44 (2012),
  pp.~1--26.

\bibitem{CleGarLanSchStu21}
{\sc E.~Cleary, A.~Garbuno-Inigo, S.~Lan, T.~Schneider, and A.~M. Stuart}, {\em
  Calibrate, emulate, sample}, J. Comput. Phys., 424 (2021), p.~109716.

\bibitem{CooGibGil08}
{\sc A.~R. Cook, G.~J. Gibson, and C.~A. Gilligan}, {\em Optimal observation
  times in experimental epidemic processes}, Biometrics, 64 (2008),
  pp.~860--868.

\bibitem{Couvreux21a}
{\sc F.~Couvreux, F.~Hourdin, D.~Williamson, R.~Roehrig, V.~Volodina,
  N.~Villefranque, C.~Rio, O.~Audouin, J.~Salter, E.~Bazile, et~al.}, {\em
  Process-based climate model development harnessing machine learning: {I.} {A}
  calibration tool for parameterization improvement}, J. Adv. Model. Earth
  Sys., 13 (2021), p.~e2020MS002217.

\bibitem{pre:DasStu13}
{\sc M.~Dashti and A.~M. Stuart}, {\em The {{B}ayesian} approach to inverse
  problems}, arXiv preprint arXiv:1302.6989,  (2013).

\bibitem{de-Rooy13a}
{\sc W.~C. {de Rooy}, P.~Bechtold, K.~Fr{\"o}hlich, C.~Hohenegger, H.~Jonker,
  D.~Mironov, A.~P. Siebesma, J.~Teixeira, and J.-I. Yano}, {\em Entrainment
  and detrainment in cumulus convection: an overview}, Quart. J. Roy. Meteor.
  Soc., 139 (2013), pp.~1--19.

\bibitem{DroMcGPet13}
{\sc C.~C. Drovandi, J.~M. McGree, and A.~N. Pettitt}, {\em Sequential monte
  carlo for {B}ayesian sequentially designed experiments for discrete data},
  Comput. Stat. Data Anal., 57 (2013), pp.~320--335.

\bibitem{Dunbar21a}
{\sc O.~R.~A. Dunbar, A.~Garbuno-Inigo, T.~Schneider, and A.~M. Stuart}, {\em
  Calibration and uncertainty quantification of convective parameters in an
  idealized {GCM}}, J. Adv. Model. Earth Sys., 13 (2021), p.~e2020MS002454.

\bibitem{duncan2021ensemble}
{\sc A.~B. Duncan, A.~M. Stuart, and M.-T. Wolfram}, {\em Ensemble inference
  methods for models with noisy and expensive likelihoods}, arXiv preprint
  arXiv:2104.03384,  (2021).

\bibitem{emanuel1995report}
{\sc K.~Emanuel, D.~Raymond, A.~Betts, L.~Bosart, C.~Bretherton,
  K.~Droegemeier, B.~Farrell, J.~M. Fritsch, R.~Houze, M.~Le~Mone, et~al.},
  {\em Report of the first prospectus development team of the us weather
  research program to noaa and the nsf}, Bull. Am. Meteorol. Soc.,  (1995),
  pp.~1194--1208.

\bibitem{EmeRey13}
{\sc A.~A. Emerick and A.~C. Reynolds}, {\em Ensemble smoother with multiple
  data assimilation}, Computers \& Geosciences, 55 (2013), pp.~3--15.

\bibitem{book:FedHac97}
{\sc V.~V. Fedorov and P.~Hackl}, {\em Model-oriented design of experiments},
  vol.~125, Springer Science \& Business Media, 1997.

\bibitem{Frierson07b}
{\sc D.~M.~W. Frierson}, {\em The dynamics of idealized convection schemes and
  their effect on the zonally averaged tropical circulation}, J. Atmos. Sci.,
  64 (2007), pp.~1959--1976.

\bibitem{Frierson06a}
{\sc D.~M.~W. Frierson, I.~M. Held, and P.~Zurita-Gotor}, {\em A gray-radiation
  aquaplanet moist {GCM}. {Part~I}: Static stability and eddy scale}, J. Atmos.
  Sci., 63 (2006), pp.~2548--2566.

\bibitem{GCSS93}
{\sc {GEWEX Cloud System Science Team}}, {\em The {GEWEX} {C}loud {S}ystem
  {S}tudy ({GCSS})}, Bull. Am. Meteorol. Soc., 74 (1993), pp.~387 -- 400.

\bibitem{Gey11}
{\sc C.~J. Geyer}, {\em Introduction to {M}arkov {C}hain {M}onte {C}arlo}, in
  Handbook of {M}arkov {C}hain {M}onte {C}arlo, S.~Brooks, A.~Gelman, G.~L.
  Jones, and X.-L. Meng, eds., Handbooks of {M}odern {S}tatistical {M}ethods,
  Chapman and Hall/CRC, 2011, ch.~1, pp.~3--48.

\bibitem{Hohenegger11a}
{\sc C.~Hohenegger and C.~S. Bretherton}, {\em Simulating deep convection with
  a shallow convection scheme}, Atmos. Chem. Phys., 11 (2011),
  pp.~10389--10406.

\bibitem{Hourdin21b}
{\sc F.~Hourdin, D.~Williamson, C.~Rio, F.~Couvreux, R.~Roehrig,
  N.~Villefranque, I.~Musat, F.~B. Diallo, L.~Fairhead, and V.~Volodina}, {\em
  Process-based climate model development harnessing machine learning: {II.}
  model calibration from single column to global}, J. Adv. Model. Earth Sys.,
  13 (2021), p.~e2020MS002225.

\bibitem{DunHowSch22}
{\sc M.~F. Howland, O.~R.~A. Dunbar, and T.~Schneider}, {\em Parameter
  uncertainty quantification in an idealized {GCM} with a seasonal cycle}, J.
  Adv. Model. Earth Sys., 14 (2022), p.~e2021MS002735.

\bibitem{HuaMar14}
{\sc X.~Huan and Y.~Marzouk}, {\em Gradient-based stochastic optimization
  methods in {B}ayesian experimental design}, Int. J. Uncertain. Quantif., 4
  (2014).

\bibitem{HuaMar13}
{\sc X.~Huan and Y.~M. Marzouk}, {\em Simulation-based optimal {B}ayesian
  experimental design for nonlinear systems}, J. Comput. Phys., 232 (2013).

\bibitem{IglLawStu13}
{\sc M.~A. Iglesias, K.~J. Law, and A.~M. Stuart}, {\em Ensemble {K}alman
  methods for inverse problems}, Inverse Problems, 29 (2013), p.~045001.

\bibitem{book:KaiSom06}
{\sc J.~Kaipio and E.~Somersalo}, {\em Statistical and computational inverse
  problems}, vol.~160, Springer Science \& Business Media, 2006.

\bibitem{Kalnay03a}
{\sc E.~Kalnay}, {\em Atmospheric Modeling, Data Assimilation and
  Predictability}, Cambridge Univ. Press, Cambridge, UK, 2003.

\bibitem{Kaspi11a}
{\sc Y.~Kaspi and T.~Schneider}, {\em Winter cold of eastern continental
  boundaries induced by warm ocean waters}, Nature, 471 (2011), pp.~621--624.

\bibitem{Kaspi13c}
\leavevmode\vrule height 2pt depth -1.6pt width 23pt, {\em The role of
  stationary eddies in shaping midlatitude storm tracks}, J. Atmos. Sci., 70
  (2013), pp.~2596--2613.

\bibitem{KenOHa00}
{\sc M.~C. Kennedy and A.~O'Hagan}, {\em Predicting the output from a complex
  computer code when fast approximations are available}, Biometrika, 87 (2000),
  pp.~1--13.

\bibitem{Kennedy01a}
\leavevmode\vrule height 2pt depth -1.6pt width 23pt, {\em Bayesian calibration
  of computer models}, J. Roy. Statist. Soc. B, 63 (2001), pp.~425--464.

\bibitem{Khairoutdinov09a}
{\sc M.~F. Khairoutdinov, S.~K. Krueger, C.-H. Moeng, P.~A. Bogenschutz, and
  D.~A. Randall}, {\em Large-eddy simulation of maritime deep tropical
  convection}, J. Adv. Model. Earth Sys., 1 (2009), pp.~Art. \#15, 13 pp.

\bibitem{KimLuMyuPitSte14}
{\sc W.~Kim, M.~A. Pitt, Z.-L. Lu, M.~Steyvers, and J.~I. Myung}, {\em A
  hierarchical adaptive approach to optimal experimental design}, Neural
  Comput., 26 (2014), pp.~2465--2492.

\bibitem{LevStu21-pre}
{\sc M.~E. Levine and A.~M. Stuart}, {\em A framework for machine learning of
  model error in dynamical systems}, arXiv preprint arxiv:2107.06658,  (2021).

\bibitem{Levine15a}
{\sc X.~Levine and T.~Schneider}, {\em Baroclinic eddies and the extent of the
  {H}adley circulation: An idealized {GCM} study}, J. Atmos. Sci., 72 (2015),
  pp.~2744--2761.

\bibitem{FoxLi17}
{\sc Q.~Li and B.~Fox-Kemper}, {\em Assessing the effects of langmuir
  turbulence on the entrainment buoyancy flux in the ocean surface boundary
  layer}, J. Phys. Oceanogr., 47 (2017), pp.~2863--2886.

\bibitem{Liu01a}
{\sc C.~Liu, M.~W. Moncrieff, and W.~W. Grabowski}, {\em Hierarchical modelling
  of tropical convective systems using explicit and parametrized approaches},
  Quart. J. Roy. Meteor. Soc., 127 (2001), pp.~493--515.

\bibitem{LonScaTemWan13}
{\sc Q.~Long, M.~Scavino, R.~Tempone, and S.~Wang}, {\em Fast estimation of
  expected information gains for {B}ayesian experimental designs based on
  {L}aplace approximations}, Comput. Methods Appl. Mech. Eng., 259 (2013),
  pp.~24--39.

\bibitem{Lopez-Gomez22a}
{\sc I.~Lopez-Gomez, C.~Christopoulos, H.~L. Ervik, O.~R.~A. Dunbar, Y.~Cohen,
  and T.~Schneider}, {\em Training physics-based machine-learning
  parameterizations with gradient-free ensemble {K}alman methods}, J. Adv.
  Model. Earth Sys.,  (2022).
\newblock In review.

\bibitem{Lorenz98a}
{\sc E.~N. Lorenz and K.~A. Emanuel}, {\em Optimal sites for supplementary
  weather observations: Simulation with a small model}, J. Atmos. Sci., 55
  (1998), pp.~399--414.

\bibitem{Matheou14a}
{\sc G.~Matheou and D.~Chung}, {\em Large-eddy simulation of stratified
  turbulence. {Part~II}: Application of the stretched-vortex model to the
  atmospheric boundary layer}, J. Atmos. Sci., 71 (2014), pp.~4439--4460.

\bibitem{Merlis11a}
{\sc T.~M. Merlis and T.~Schneider}, {\em Changes in zonal surface temperature
  gradients and walker circulations in a wide range of climates}, J. Climate,
  24 (2011), pp.~4757--4768.

\bibitem{book:NotSanWil18}
{\sc W.~I. Notz, T.~J. Santner, and B.~J. Williams}, {\em The design and
  analysis of computer experiments}, Springer Series in Statistics, Springer,
  2nd ed.~ed., 2018.

\bibitem{OGorman11a}
{\sc P.~A. O'Gorman}, {\em The effective static stability experienced by eddies
  in a moist atmosphere}, J. Atmos. Sci., 68 (2011), pp.~75--90.

\bibitem{OGorman11b}
{\sc P.~A. O'Gorman, N.~Lamquin, T.~Schneider, and M.~S. Singh}, {\em The
  relative humidity in an isentropic advection--condensation model: Limited
  poleward influence and properties of subtropical minima}, J. Atmos. Sci., 68
  (2011), pp.~3079--3093.

\bibitem{OGorman08c}
{\sc P.~A. O'Gorman and T.~Schneider}, {\em Energy of midlatitude transient
  eddies in idealized simulations of changed climates}, J. Climate, 21 (2008),
  pp.~5797--5806.

\bibitem{OGorman08b}
\leavevmode\vrule height 2pt depth -1.6pt width 23pt, {\em The hydrological
  cycle over a wide range of climates simulated with an idealized {GCM}}, J.
  Climate, 21 (2008), pp.~3815--3832.

\bibitem{OGorman09b}
\leavevmode\vrule height 2pt depth -1.6pt width 23pt, {\em The physical basis
  for increases in precipitation extremes in simulations of 21st-century
  climate change}, Proc. Natl. Acad. Sci., 106 (2009), pp.~14773--14777.

\bibitem{OGorman09a}
\leavevmode\vrule height 2pt depth -1.6pt width 23pt, {\em Scaling of
  precipitation extremes over a wide range of climates simulated with an
  idealized {GCM}}, J. Climate, 22 (2009), pp.~5676--5685.

\bibitem{Oliver08a}
{\sc D.~S. Oliver, A.~C. Reynolds, and N.~Liu}, {\em Inverse Theory for
  Petroleum Reservoir Characterization and History Matching}, Cambridge Univ.
  Press, 2008.

\bibitem{EidKarPag20}
{\sc J.~Paglia, J.~Eidsvik, and J.~Karvanen}, {\em Efficient spatial designs
  using {H}ausdorff distances and {B}ayesian optimisation}, Statistical
  modeling for safer drilling operations,  (2020), p.~77.

\bibitem{Pan05}
{\sc L.~Paninski}, {\em Asymptotic theory of information-theoretic experimental
  design}, Neural Comput., 17 (2005), pp.~1480--1507.

\bibitem{Pressel15a}
{\sc K.~G. Pressel, C.~M. Kaul, T.~Schneider, Z.~Tan, and S.~Mishra}, {\em
  Large-eddy simulation in an anelastic framework with closed water and entropy
  balances}, J. Adv. Model. Earth Sys., 7 (2015), pp.~1425--1456.

\bibitem{Pressel17a}
{\sc K.~G. Pressel, S.~Mishra, T.~Schneider, C.~M. Kaul, and Z.~Tan}, {\em
  Numerics and subgrid-scale modeling in large eddy simulations of
  stratocumulus clouds}, J. Adv. Model. Earth Sys., 9 (2017), pp.~1342--1365.

\bibitem{Rauber07a}
{\sc R.~M. Rauber, B.~Stevens, H.~T. Ochs, C.~Knight, B.~Albrecht, A.~Blyth,
  C.~Fairall, J.~Jensen, S.~Lasher-Trapp, O.~Mayol-Bracero, et~al.}, {\em Rain
  in shallow cumulus over ocean: {T}he {RICO} campaign}, Bull. Amer. Meteor.
  Soc., 88 (2007), pp.~1912--1928.

\bibitem{Rei11}
{\sc S.~Reich}, {\em A dynamical systems framework for intermittent data
  assimilation}, BIT Numer. Math., 51 (2011), pp.~235--249.

\bibitem{Romps16a}
{\sc D.~M. Romps}, {\em The {S}tochastic {P}arcel {M}odel: A deterministic
  parameterization of stochastically entraining convection}, J. Adv. Model.
  Earth Sys., 8 (2016), pp.~319--344.

\bibitem{ChoMarRue09}
{\sc H.~Rue, S.~Martino, and N.~Chopin}, {\em Approximate {B}ayesian inference
  for latent {G}aussian models by using integrated nested {L}aplace
  approximations}, J. R. Stat. Soc. Ser. B Methodol., 71 (2009), pp.~319--392.

\bibitem{DroMcGPetRya16}
{\sc E.~G. Ryan, C.~C. Drovandi, J.~M. McGree, and A.~N. Pettitt}, {\em A
  review of modern computational algorithms for {B}ayesian optimal design},
  Int. Stat. Rev., 84 (2016), pp.~128--154.

\bibitem{DroPetRyaTho14}
{\sc E.~G. Ryan, C.~C. Drovandi, M.~H. Thompson, and A.~N. Pettitt}, {\em
  Towards {B}ayesian experimental design for nonlinear models that require a
  large number of sampling times}, Comput. Stat. Data Anal., 70 (2014), pp.~45
  -- 60.

\bibitem{Schalkwijk15a}
{\sc J.~Schalkwijk, H.~J.~J. Jonker, A.~P. Siebesma, and E.~{Van Meijgaard}},
  {\em Weather forecasting using {GPU}-based large-eddy simulations}, Bull.
  Amer. Meteor. Soc., 96 (2015), pp.~715--723.

\bibitem{SchStu17}
{\sc C.~Schillings and A.~M. Stuart}, {\em Analysis of the ensemble {K}alman
  filter for inverse problems}, SIAM J. Numer. Anal., 55 (2017),
  pp.~1264--1290.

\bibitem{Schneider99a}
{\sc T.~Schneider and S.~M. Griffies}, {\em A conceptual framework for
  predictability studies}, J. Climate, 12 (1999), pp.~3133--3155.

\bibitem{Schneider17c}
{\sc T.~Schneider, S.~Lan, A.~Stuart, and J.~Teixeira}, {\em Earth system
  modeling 2.0: A blueprint for models that learn from observations and
  targeted high-resolution simulations}, Geophys. Res. Lett., 44 (2017),
  pp.~12396--12417.

\bibitem{Schneider08c}
{\sc T.~Schneider and P.~A. O'Gorman}, {\em Moist convection and the thermal
  stratification of the extratropical troposphere}, J. Atmos. Sci., 65 (2008),
  pp.~3571--3583.

\bibitem{Schneider10a}
{\sc T.~Schneider, P.~A. O'Gorman, and X.~J. Levine}, {\em Water vapor and the
  dynamics of climate changes}, Rev. Geophys., 48 (2010), p.~RG3001.
\newblock doi:10.1029/2009RG000302.

\bibitem{Schneider22v}
{\sc T.~Schneider, A.~M. Stuart, and J.~Wu}, {\em Ensemble {K}alman inversion
  for sparse learning of dynamical systems from time-averaged data}, J. Comp.
  Phys.,  (2021).

\bibitem{Shen20a}
{\sc Z.~Shen, K.~G. Pressel, Z.~Tan, and T.~Schneider}, {\em Statistically
  steady state large-eddy simulations forced by an idealized {GCM}: 1. forcing
  framework and simulation characteristics}, J. Adv. Model. Earth Sys., 12
  (2020), p.~e2019MS001814.

\bibitem{JarSchSheSriZhi21-pre}
{\sc Z.~Shen, A.~Sridhar, Z.~Tan, A.~Jaruga, and T.~Schneider}, {\em A library
  of large-eddy simulations for calibrating cloud parameterizations},
  https://essoar.org,  (2021).

\bibitem{Siebesma03}
{\sc A.~P. Siebesma, C.~S. Bretherton, A.~Brown, A.~Chlond, J.~Cuxart, P.~G.
  Duynkerke, H.~Jiang, M.~Khairoutdinov, D.~Lewellen, C.~H. Moeng, E.~Sanchez,
  B.~Stevens, and D.~E. Stevens}, {\em A large eddy simulation intercomparison
  study of shallow cumulus convection}, J. Atmos. Sci., 60 (2003),
  pp.~1201--1219.

\bibitem{Siebesma07}
{\sc A.~P. Siebesma, P.~M.~M. Soares, and J.~Teixeira}, {\em A combined
  eddy-diffusivity mass-flux approach for the convective boundary layer}, J.
  Atmos. Sci., 64 (2007), pp.~1230--1248.

\bibitem{Simmons81}
{\sc A.~J. Simmons and D.~M. Burridge}, {\em An energy and angular-momentum
  conserving vertical finite-difference scheme and hybrid vertical
  coordinates}, Mon. Wea. Rev., 109 (1981), pp.~758--766.

\bibitem{Smalley19a}
{\sc M.~Smalley, K.~Suselj, M.~Lebsock, and J.~Teixeira1}, {\em A novel
  framework for evaluating and improving parameterized subtropical marine
  boundary layer cloudiness}, Mon. Wea. Rev., 147 (2019), pp.~3241--3260.

\bibitem{Sou-etal20}
{\sc A.~N. Souza, G.~L. Wagner, A.~Ramadhan, B.~Allen, V.~Churavy, J.~Schloss,
  J.~Campin, C.~Hill, A.~Edelman, J.~Marshall, G.~Flierl, and R.~Ferrari}, {\em
  Uncertainty quantification of ocean parameterizations: Application to the
  {K}-{P}rofile-{P}arameterization for penetrative convection}, J. Adv. Model.
  Earth Sys., 12 (2020), p.~e2020MS002108.

\bibitem{Stephens05}
{\sc G.~L. Stephens}, {\em Cloud feedbacks in the climate system: A critical
  review}, J. Climate, 18 (2005), pp.~237--273.

\bibitem{Ste-etal03}
{\sc B.~Stevens, D.~H. Lenschow, G.~Vali, H.~Gerber, A.~Bandy, B.~Blomquist,
  J.~L. Brenguier, C.~S. Bretherton, F.~Burnet, T.~Campos, S.~Chai, I.~Faloona,
  D.~Friesen, S.~Haimov, K.~Laursen, D.~K. Lilly, S.~M. Loehrer, S.~P.
  Malinowski, B.~Morley, M.~D. Petters, D.~C. Rogers, L.~Russell,
  V.~Savic-Jovcic, J.~R. Snider, D.~Straub, M.~J. Szumowski, H.~Takagi, D.~C.
  Thornton, M.~Tschudi, C.~Twohy, M.~Wetzel, and M.~C. van Zanten}, {\em
  Dynamics and chemistry of marine stratocumulus-{DYCOMS}-{II}}, Bull. Amer.
  Meteor. Soc., 84 (2003), pp.~579 -- 594.

\bibitem{Stevens05a}
{\sc B.~Stevens, C.-H. Moeng, A.~S. Ackerman, C.~S. Bretherton, A.~Chlond,
  S.~{de Roode}, J.~Edwards, J.-C. Golaz, H.~Jiang, M.~Khairoutdinov, M.~O.
  Kirkpatrick, D.~C. Lewellen, A.~Lock, F.~M{\"u}ller, D.~E. Stevens,
  E.~Whelan, and P.~Zhu}, {\em Evaluation of large-eddy simulations via
  observations of nocturnal marine stratocumulus}, Mon. Wea. Rev., 133 (2005),
  pp.~1443--1462.

\bibitem{SchSto94}
{\sc G.~M. Stokes and S.~E. Schwartz}, {\em The atmospheric radiation
  measurement (arm) program: Programmatic background and design of the cloud
  and radiation test bed}, Bulletin of the American Meteorological Society, 75
  (1994), pp.~1201 -- 1222.

\bibitem{Stu10}
{\sc A.~M. Stuart}, {\em Inverse problems: a {B}ayesian perspective}, Acta
  Numerica, 19 (2010), pp.~451--559.

\bibitem{Tan18a}
{\sc Z.~Tan, C.~M. Kaul, K.~G. Pressel, Y.~Cohen, T.~Schneider, and
  J.~Teixeira}, {\em An extended eddy-diffusivity mass-flux scheme for unified
  representation of subgrid-scale turbulence and convection}, J. Adv. Model.
  Earth Sys., 10 (2018), pp.~770--800.

\bibitem{book:Tar05}
{\sc A.~Tarantola}, {\em Inverse problem theory and methods for model parameter
  estimation}, vol.~89, {SIAM}, 2005.

\bibitem{GahHajTsi17}
{\sc P.~Tsilifis, R.~G. Ghanem, and P.~Hajali}, {\em Efficient {B}ayesian
  experimentation using an expected information gain lower bound}, SIAM-ASA J.
  Uncertain., 5 (2017), pp.~30--62.

\bibitem{Uci00}
{\sc D.~Uci{\'n}ski}, {\em Optimal selection of measurement locations for
  parameter estimation in distributed processes}, Int. J. Appl. Math. Comput.
  Sci, 10 (2000), pp.~357--379.

\bibitem{FedUci07}
{\sc D.~Uci{\'n}ski and M.~Patan}, {\em {D}-optimal design of a monitoring
  network for parameter estimation of distributed systems}, J. Glob. Optim., 39
  (2007), pp.~291--322.

\bibitem{BraVan01}
{\sc M.~{van de Wal} and B.~{de Jager}}, {\em A review of methods for
  input/output selection}, Automatica, 37 (2001), pp.~487--510.

\bibitem{Vial13a}
{\sc J.~Vial, J.-L. Dufresne, and S.~Bony}, {\em On the interpretation of
  inter-model spread in {CMIP5} climate sensitivity estimates}, Clim. Dyn., 41
  (2013), pp.~3339--3362.

\bibitem{Webb13b}
{\sc M.~J. Webb, F.~H. Lambert, and J.~M. Gregory}, {\em Origins of differences
  in climate sensitivity, forcing and feedback in climate models}, Clim. Dyn.,
  40 (2013), pp.~677--707.

\bibitem{Wei18a}
{\sc H.-H. Wei and S.~Bordoni}, {\em Energetic constraints on the {ITCZ}
  position in idealized simulations with a seasonal cycle}, J. Adv. Model.
  Earth Sys., 10 (2018).

\bibitem{book:RasWil06}
{\sc C.~K. Williams and C.~E. Rasmussen}, {\em Gaussian processes for machine
  learning}, vol.~2, MIT press Cambridge, MA, 2006.

\bibitem{Williams11a}
{\sc P.~D. Williams}, {\em The {RAW} filter: An improvement to the
  {R}obert--{A}sselin filter in semi-implicit integrations}, Mon. Wea. Rev.,
  139 (2011), pp.~1996--2007.

\bibitem{Wills17a}
{\sc R.~C. Wills, X.~J. Levine, and T.~Schneider}, {\em Local energetic
  constraints on {W}alker circulation strength}, J. Atmos. Sci., 74 (2017),
  pp.~1907--1922.

\bibitem{BuiCheEmaMagSunZha19}
{\sc F.~Zhang, Y.~Q. Sun, L.~Magnusson, R.~Buizza, S.-J. Lin, J.-H. Chen, and
  K.~Emanuel}, {\em What is the predictability limit of midlatitude weather?},
  J. Atmos. Sci., 76 (2019), pp.~1077 -- 1091.

\bibitem{Zhang13a}
{\sc M.~Zhang, C.~S. Bretherton, P.~N. Blossey, P.~H. Austin, J.~T. Bacmeister,
  S.~Bony, F.~Brient, S.~K. Cheedela, A.~Cheng, A.~D. {Del Genio}, S.~R. {de
  Roode}, et~al.}, {\em {CGILS:} {R}esults from the first phase of an
  international project to understand the physical mechanisms of low cloud
  feedbacks in general circulation models}, J. Adv. Model. Earth Sys., 5
  (2013), pp.~826--842.

\end{thebibliography}

\appendix

\section{Calibrate-Emulate-Sample with design}\label{sec:CES}

\begin{figure}[h]
 \centering
  \includegraphics[width=\textwidth]{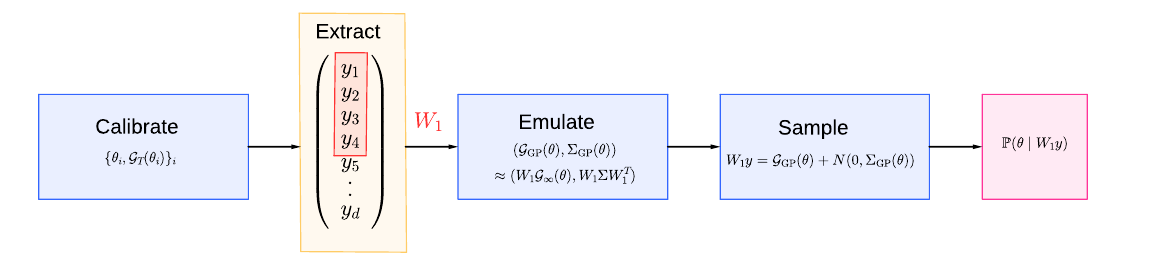}
 \caption{Procedure of the uncertainty quantification framework (blue), to produce output (pink). A restriction operator $W_1$ extracting a subset of the GCM output (yellow); the subsequent emulate and sample stages may be performed in parallel for all $W_i$, from a single calibration run.  }
\label{fig:CES_des}
\end{figure}

Key to the success of this work, is the ability to efficiently calculate the the posterior distribution (in particular the covariance), which is needed to calculate the utility function \eqref{eq:Dutil} at all designs. We present a methodology: calibrate-extract-emulate-sample, (CEES) which allows for the parallel sampling of the posterior distribution at all designs with a combined total of $\mmm[O]$(100) evaluations of our forward model.

The methodology is based on the calibrate-emulate-sample (CES) algorithm, for full details of the individual stages see \cite{CleGarLanSchStu21,Dunbar21a}, here we present an overview and motivation. The core purpose of CES is to form a computationally cheap statistical emulator of $\mmm[G]_\infty$ from intelligently chosen samples of $\mmm[G]_T$; then one is able to solve the Bayesian inverse problem for the emulated $\mmm[G]_{\infty}$ with a sampling method. We achieve this by using Gaussian process emulators, trained on the samples of the (noisy and expensive) forward map. The Gaussian process mean function is naturally smoother than the data it is trained on \cite{Kennedy01a,book:NotSanWil18}, and is capable of representing the the noise of the forward model within the covariance function, leading to a smooth likelihood function that is quick to evaluate. The training points for the Gaussian Process are given by applying an optimization scheme, EKI (Ensemble Kalman Inversion), \cite{CheOli12,IglLawStu13,SchStu17} to the inverse problem in its finite-time averaged form \eqref{eq:globalIPfin}. Theoretical work shows that noisy continuous-time versions of EKI exhibit an averaging effect that skips over 
fluctuations superimposed onto the ergodic averaged forward model \cite{duncan2021ensemble}, and similar effects are observed in practice for EKI, thus it is highly suited to optimization of parameters coming from a noisy, expensive model without derivatives available. Ensemble Kalman methods are scalable to very high dimensional problems \cite{Kalnay03a,Oliver08a} with use of localization and regularization.

Let $D$ index a finite space of designs. Given a time $T>0$, and prior on $\bbb[\theta]$ with prior mean $\bbb[\theta]^*$. Draw a sample  $\bbb[y] = \mmm[G]_T(\bbb[\theta]^*,\bbb[v]^{(0)})$, from initial condition $\bbb[v]^{(0)}$: 
\begin{enumerate}
 \item \textbf{Calibrate:} We solve \eqref{eq:globalIPfin} with $\bbb[y]$ using evaluations of $\mmm[G]_T$ in an optimization sense, where we minimize the functional. 
 \begin{equation}\label{eq:obj_eki}
 \Phi_{T}(\bbb[\theta],\bbb[y]) = \| \bbb[y] - \mmm[G]_T(\bbb[\theta])\|^2_{2\Sigma}. 
 \end{equation}
 The notation $\| \cdot \|_{\Sigma} = \|\Sigma^{-\frac{1}{2}} \cdot\|_2$ is the Mahalanobis distance. We drop the notation of the initial conditions, which are drawn at random from the invariant distribution for every evaluation of $\mmm[G]_T$. The weight $2\Sigma$ is the sum of internal variability of $\mmm[G]_T$ and of $\bbb[y]$. The optimization is performed using several iterations the Ensemble Kalman Inversion algorithm. This leads to $\{\bbb[\theta]_i,\mmm[G]_j(\bbb[\theta]_j)\}_{j=1}^{J}$ of input-output pairs that are localized around the optimal parameter value.
 \item \textbf{Extract:} For each design $k\in D$, we apply the restriction mapping $W_k$ to the forward map,  $\{\bbb[\theta]_j,W_k\mmm[G]_T(\bbb[\theta]_j)\}_{j=1}^{J}$, and apply the following \textbf{Emulate(k)} and \textbf{Sample(k)} stages.
 \item \textbf{Emulate(k):} We decorrelate the data space with an SVD on the internal variability covariance $\Sigma$, yielding a change-of-basis matrix $V$. We train Gaussian process emulators, on the pairs $\{\bbb[\theta]_j,VW_k\mmm[G]_T(\bbb[\theta]_j)\}_{j=1}^{J}$, yielding $(\mmm[G]_{\text{GP}}(\bbb[\theta]), \Sigma_{\text{GP}}(\bbb[\theta]))$, where $\mmm[G]_{\text{GP}}\approx VW_k\mmm[G]_\infty(\bbb[\theta])$ (crucially $\mmm[G]_\infty$ and not $\mmm[G]_T$) and $\Sigma_{\text{GP}}(\bbb[\theta])\approx V W_k\Sigma W_k^T V^T$.
 \item \textbf{Sample(k):} We now solve the inverse problem \eqref{eq:restrIPinf}, This is feasible as the emulator provides us with an approximation of $\mmm[G]_\infty$ (not just $\mmm[G]_T$). The posterior distribution associated with \eqref{eq:restrIPinf} is proportional to a product of prior and likelihood contribution from Bayes theorem. Explicitly, for a Gaussian prior $N(\bbb[m],C)$ on the computational parameters, and the likelihood dependent on the emulator, we write the MCMC objective function (also known as the log-posterior) as
 \begin{align*}
 \Phi_{\mathrm{MCMC}}(\bbb[\theta],VW_k\bbb[y]) =& \frac{1}{2}\| V W_k\bbb[y] - \mathcal{G}_{\mathrm{GP}}(\bbb[\theta])\|^2_{\Sigma_{\mathrm{GP}}(\bbb[\theta])} + \frac{1}{2}\log \det\Sigma_{\text{GP}}(\bbb[\theta]) \\
 &+\frac{1}{2}\|\bbb[\theta] - \bbb[m] \|^2_C\,.
 \end{align*}
The posterior is then given by 
 \[
 \mathbb{P}(\bbb[\theta]\mid VW_k\bbb[y]) \propto\exp(-\Phi_{MCMC}(\bbb[\theta],VW_k \bbb[y])).
 \]
 This can be sampled with a standard random walk metropolis sampling algorithm. In practice we run the algorithm for $2\times10^5$ samples, discarding the first $10^5$ as spin-up.
 
\end{enumerate}
The CEES algorithm is illustrated in Figure~\ref{fig:CES_des}. We then collect the posterior distributions $\{\bbb[\theta]\mid W_k\bbb[y]\}_k, \ \forall k\in D$ and calculate the utility function using \eqref{eq:Dutil}. In particular the algorithm requires J model evaluations independent of the number of designs.

The CES algorithm is used to solve \eqref{eq:localIPinf} at a given design $\tilde{k}$, by calibrating with the corresponding objective function for the limited-area data, followed by emulate and sample stages at $\tilde{k}$.

\section{Results for three-latitude stencil}\label{sec:3stencil}

\begin{figure}[h]
 \centering
\includegraphics[width=0.65\textwidth]{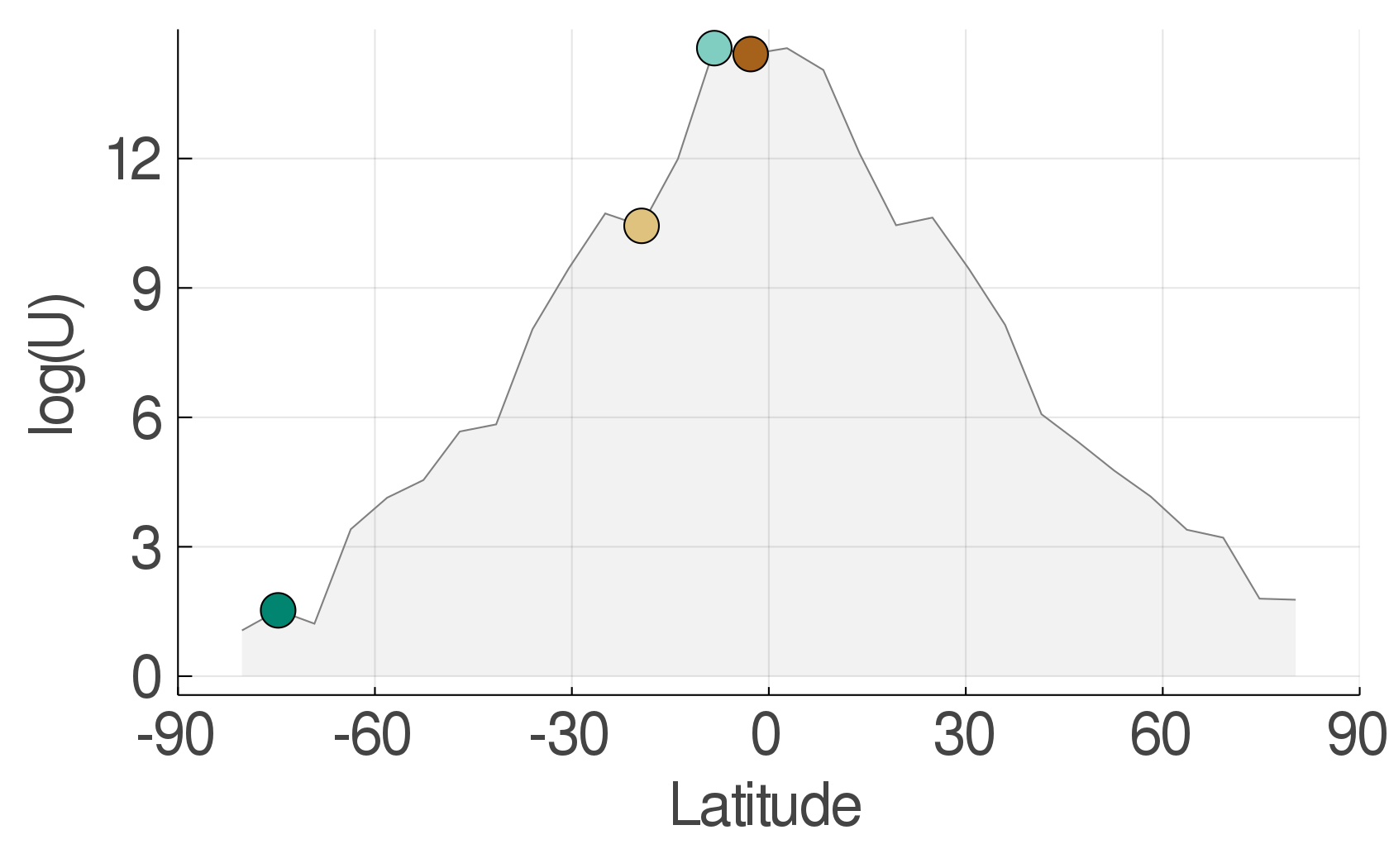}
\caption{Logarithm of the data utility as a function of latitude, with designs corresponding to a three-latitude stencil, the center of which is plotted. The colored discs signify the four representative designs, which are used in the uncertainty quantification experiment.}
\label{f:util_3stencil}
\end{figure}

\begin{figure}[h]
    \centering
    \includegraphics[width=0.8\textwidth]{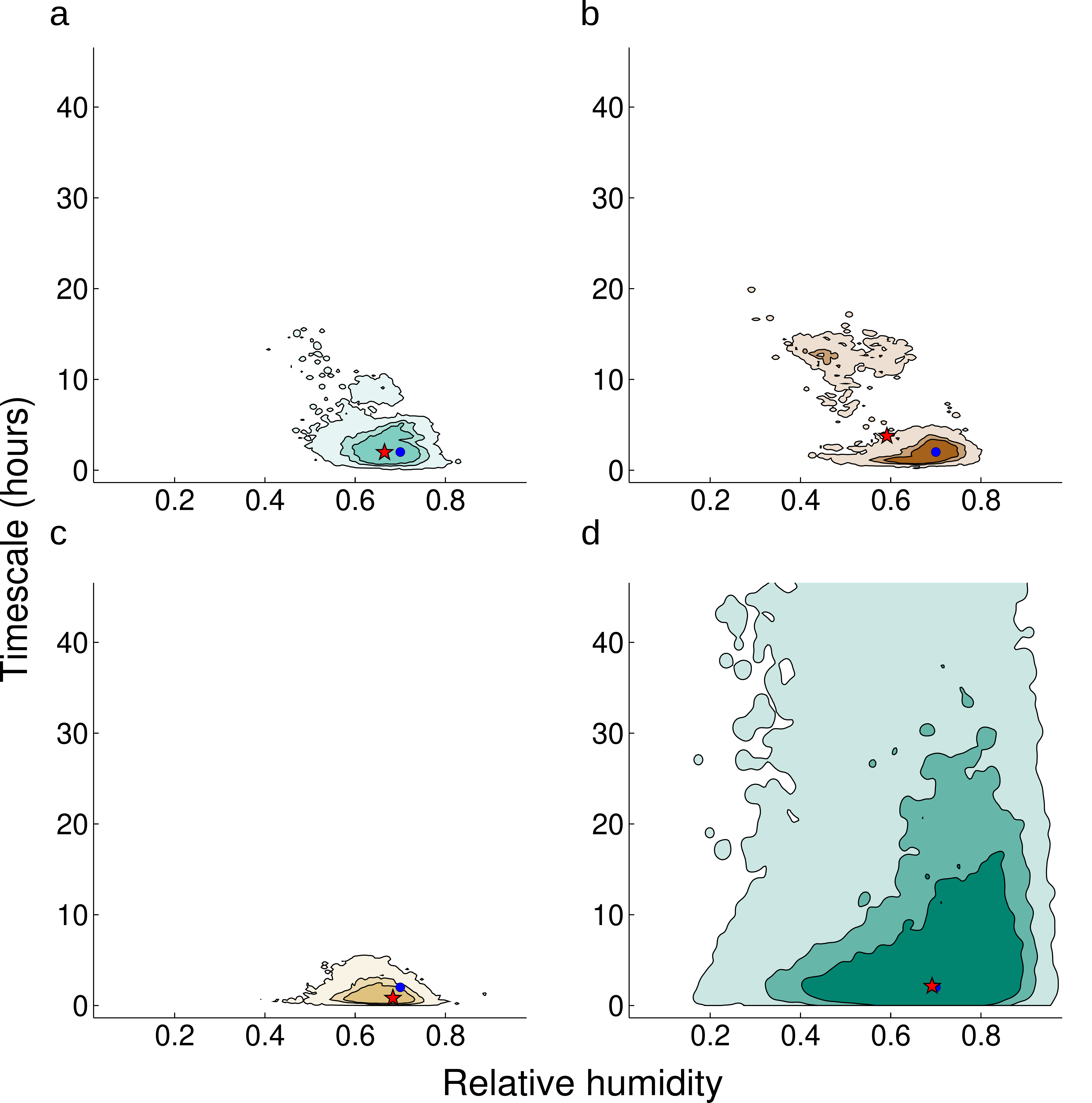}
     \caption{Posterior distributions for convection parameters learned from data restricted to different design points. The drawn contours bound 50\%, 75\% and 99\% of the distribution. Panels a---d correspond to designs -$8^\circ$, $-3^\circ$, $-19^\circ$, and $-75^\circ$), ordered as points of decreasing utility in Figure~\ref{f:util_3stencil}. The true utility of these distributions are $126.0$, $35.3$, $97.5$,  and $2.1$. The true parameter values in the control simulation are given by the blue circle. The parameters found to be optimal in the calibration scheme (given a single random realization of data) are given by the red star in each case.}
    \label{f:posterior_3stencil}
\end{figure}

For the statistically stationary case, we increase the stencil size to $\ell=3$. Here, we have 30 designs indexed from south to north poles. We plot the logarithm of the utility against the designs in Figure~\ref{f:util_3stencil}. The center of the three-latitude stencil is take as a representative latitude for that design. The colored discs represent the designs centered on latitudes -$8^\circ$, $-3^\circ$, $-19^\circ$, and $-75^\circ$,  
in decreasing order of utility on the plot. The increase in spatial extent smooths the design landscape. We validate the optimal design methodology by taking a data sample at each of these representative designs. We then apply the uncertainty quantification stage of the algorithm for each design to obtain the posterior distributions for the convection parameters given each data. The distributions are displayed in Figure~\ref{f:posterior_3stencil}; panels a---d are ordered according to decreasing predicted utility given by Figure~\ref{f:util_3stencil}. The true utilities for the distributions a---d are $126.0$, $35.3$, $97.5$, and $2.1$. In this case, the algorithm has identified the design with maximal utility centered at $-8^\circ$ (Figure~\ref{f:util_3stencil}a), where analysis of precipitation and parameterized tendencies would suggest the ITCZ region centered at $-3^\circ$ that presents the bimodal distribution (Figure~\ref{f:util_3stencil}b). 

\section{Results for 30-day time averages}\label{sec:30day}

\begin{figure}[h]
 \centering
\includegraphics[width=\textwidth]{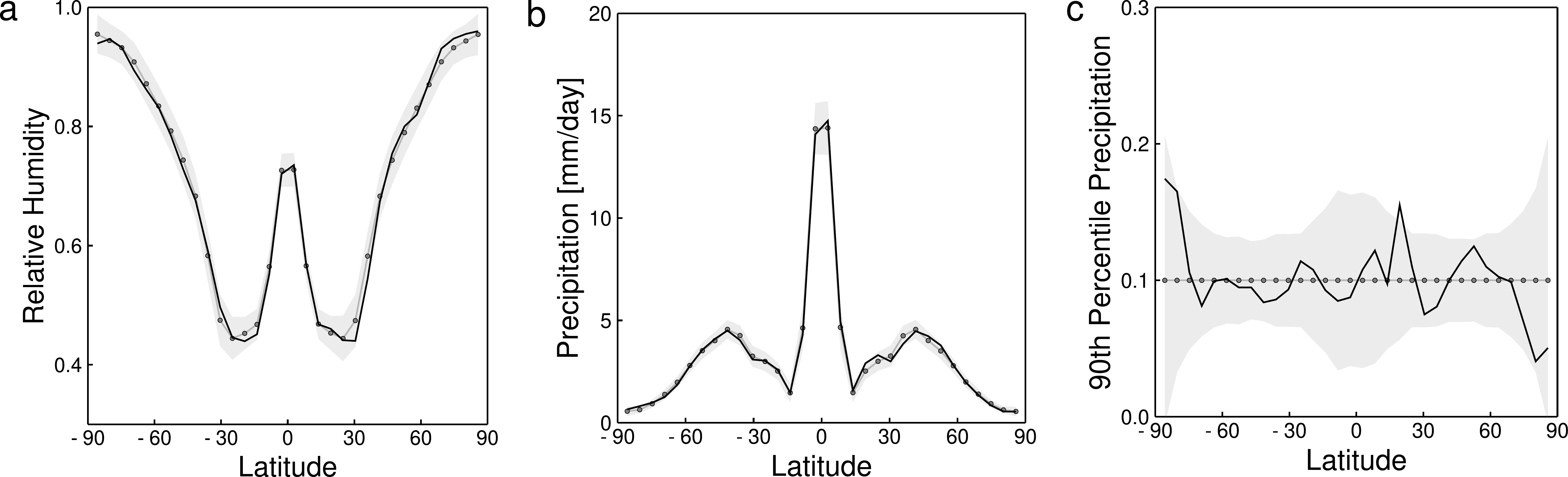}
\caption{Aggregated climate statistics  in the statistically stationary control simulation, with parameters set to the mean of the prior $\bbb[\theta]^*$. The mean (grey lines) and 95\% confidence intervals (shading) of the data are plotted against latitude. 
One realization of the data is shown (black line). No noise is added here. }
\label{30day_data_meanparam}
\end{figure}

\begin{figure}[h]
 \centering
\includegraphics[width=0.65\textwidth]{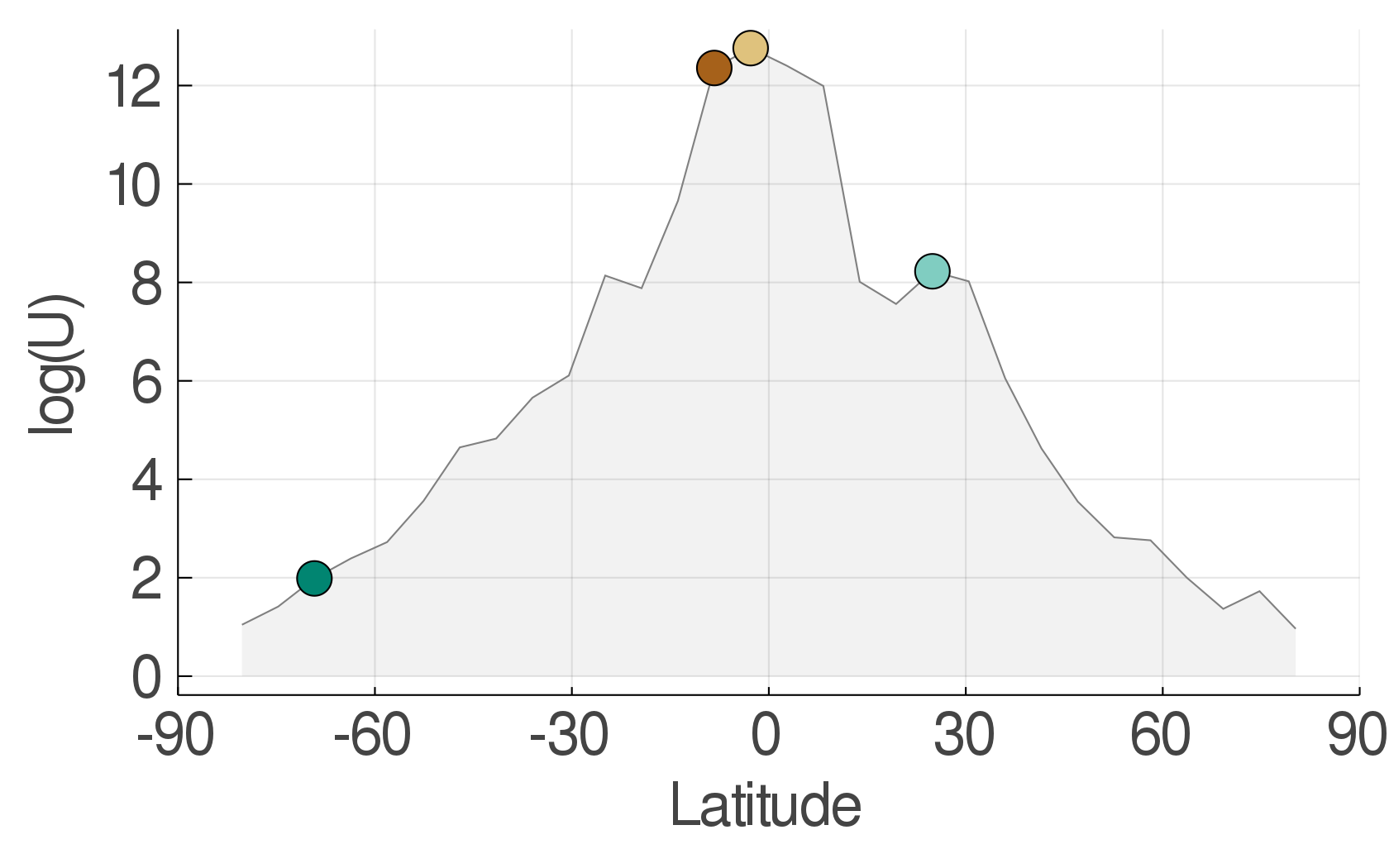}
\caption{Logarithm of the data utility as a function of latitude, with designs represented by a node at the center of each stencil (comprised of three neighboring latitudes). The colored discs signify the four representative designs indicated in Fig.~\ref{f:data_phys}, which are used in the uncertainty quantification experiment.}
\label{f:30day_util_3stencil}
\end{figure}

For the statistically stationary case, we also run a suite of experiments for time-averages of $T=30$ days. The first control simulation at the prior mean $\bbb[\theta]^*$ produces the 600 samples of 30-day averaged control statistics, with which we use to approximate $\Sigma(\theta)$. The statistics are represented in Figure~\ref{30day_data_meanparam}. We run an experiment for three-stencil designs ($\ell=3$). Here we have 30 designs are indexed from south to north poles. We plot the logarithm of the utility against the designs in Figure~\ref{f:30day_util_3stencil}. The colored discs represent the designs centred on latitudes $-3^\circ$, $-8^\circ$, $25^\circ$, and $-69^\circ$ (in decreasing order of utility on the plot). 

\begin{figure}[h]
 \centering
\includegraphics[width=\textwidth]{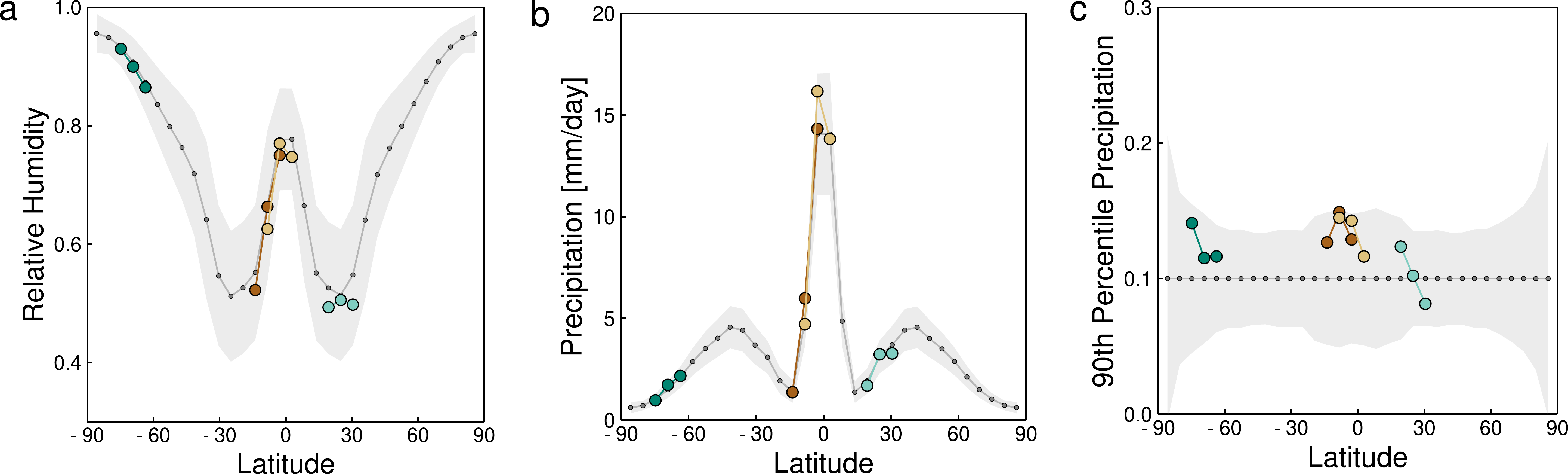}
\caption{Aggregated climate statistics in the statistically stationary control simulation using the ground truth parameters. Mean (grey lines) and 95\% confidence intervals (shading) of the data are plotted against latitude. Additional inflation noise is added. Each set of colored discs represents a 30-day realization of inflated GCM data coming from a different three-latitude design used in the experiment.}
\label{f:30day_data_phys_3stencil}
\end{figure}

\begin{figure}[h]
 \centering
\includegraphics[width=0.8\textwidth]{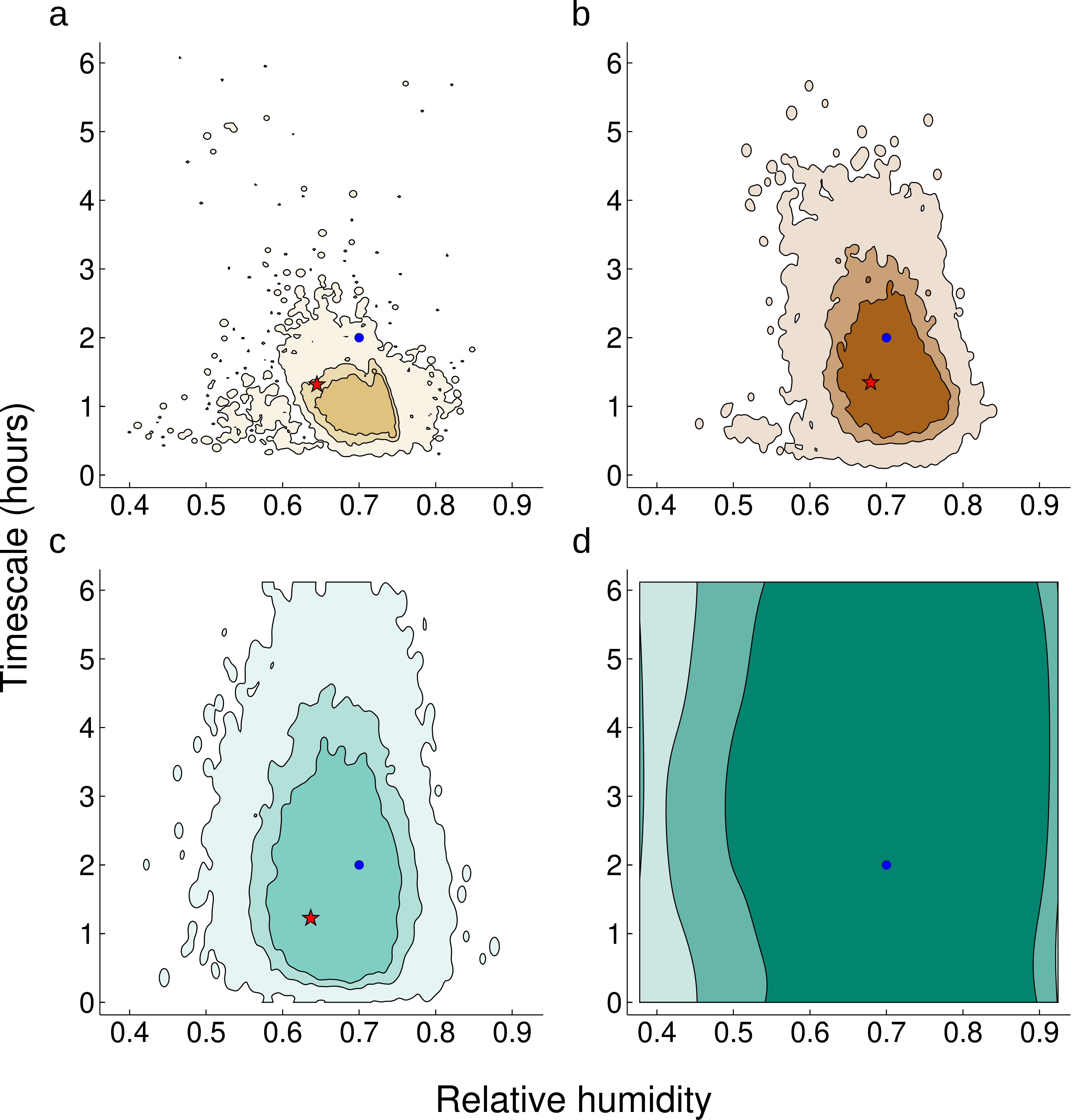}
\caption{Posterior distributions for convection parameters learned from data restricted to different design points. The drawn contours bound 50\%, 75\% and 99\% of the distribution. Panels a---d correspond to 3-stencil designs with centres at $-8^\circ$, $-3^\circ$, $25^\circ$, and $-70^\circ$, ordered to express learning from data at decreasingly informative design points (i.e., points of decreasing utility in Figure~\ref{f:30day_util_3stencil}). The true parameter values in the control simulation are given by the blue circle. The parameters found to be optimal in the calibration scheme (given a single random realization of data) are given by the red star in each case (in panel (d) this is outside the plotting region). 
}
\label{f:30day_posterior_phys}
\end{figure}

To validate the optimal design methodology, we sample the ground truth data at the designs (Figure~\ref{f:30day_data_phys_3stencil}). We then obtain the posterior distributions for the convection parameters given this data. The distributions are displayed in Figure~\ref{f:30day_posterior_phys}, with panels a---d ordered according to decreasing predicted utility given by Figure~\ref{f:30day_util_3stencil}. The uncertainty of the distributions in panels a---d gives utilities $181.6$, $97.5$, $66.6$, and $1.8$. In this case, the automated algorithm has identified the optimal stencil correctly.

\begin{figure}[h]
 \centering
\includegraphics[width=0.8\textwidth]{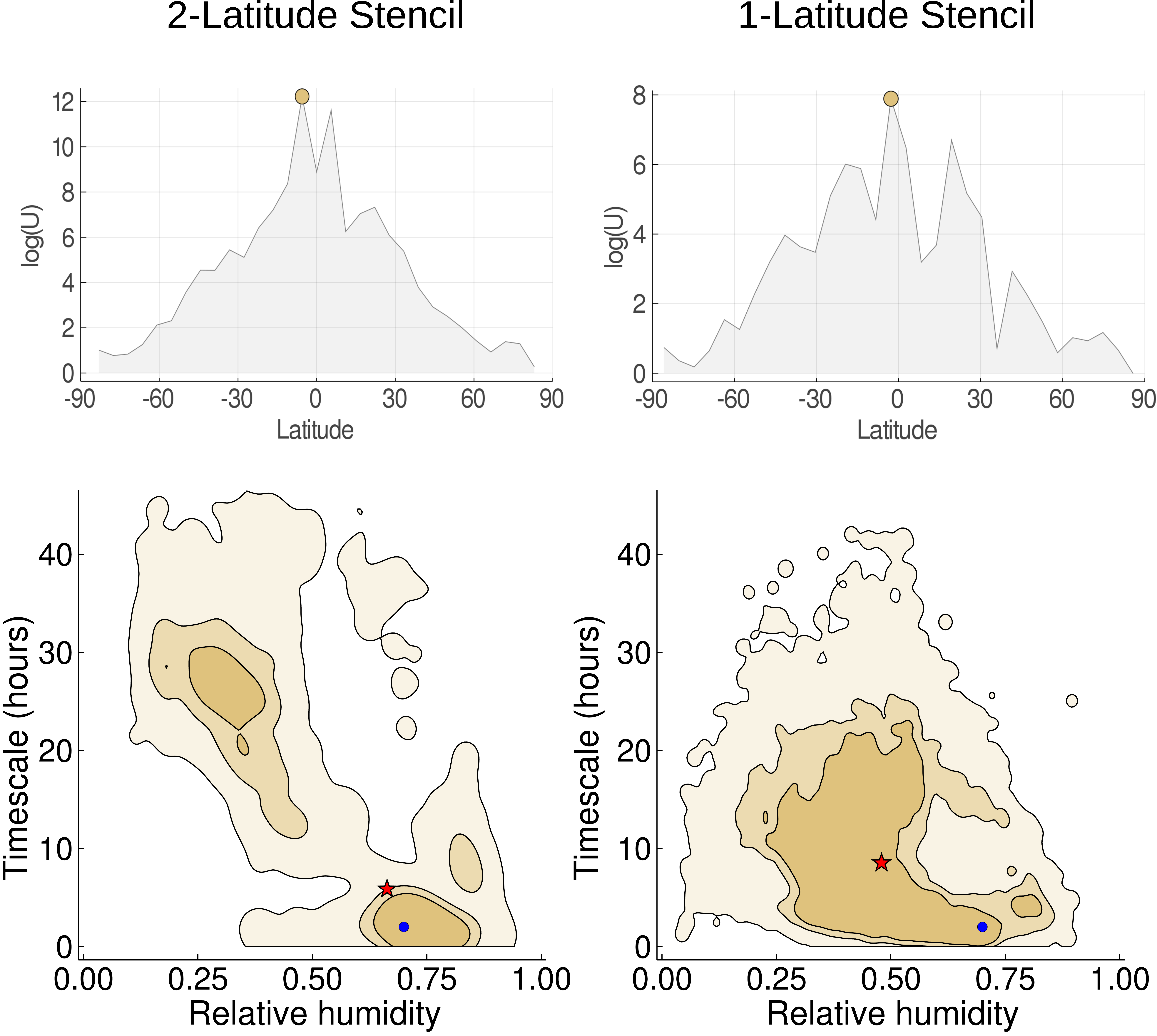}
\caption{Performance for different optimal design selections at smaller stencil sizes. The contours bound 50\%, 75\%, and 99\% of the distribution. The top row displays the logarithm of the utility plot, and the bottom row the corresponding posterior from a sample at the optimal latitude, marked by a disc at the top.}
\label{f:30day_posterior_stencils}
\end{figure}

With the choices $\ell=1$ and $2$, Figure~\ref{f:30day_posterior_stencils} shows the utility function against the latitude at the center of the stencil and the posterior distribution at the respective optimal designs. The behavior of the utility function is similar to the 90-day time averaged case. For $\ell=2$, we find optimality around the equator; for $\ell=1$, we find two additional peaks revealed at $\pm 19^\circ$ (see Figure~\ref{f:util_1stencil}). The posterior distributions are seen to be far broader than in the three-latitude case, as less information is available. The posteriors are multimodal but nevertheless capture the true parameters (blue disc) with high probability. They provide insight into the correlation structure between the parameters at the optimal design location. We observe that for these sparser designs, non-identifiability (multimodality) appears only at data from $\bbb[\theta]^\dagger$, but not at $\bbb[\theta]^*$. As a result, the optimal uncertainty is not guaranteed to be found at the location of optimal utility. This is remedied by having a better initial guess through the prior, or by having a less noisy data set from which the parameters are more identifiable. 

\end{document}